\documentclass[journal,onecolumn,font=12]{IEEEtran}
   \usepackage[pdftex]{graphicx}
\usepackage[cmex10]{amsmath}
\usepackage{pgf,tikz}
\usetikzlibrary{arrows}
\usepackage{research5IEEE}

\usepackage{amssymb}
\usepackage{color}
\usepackage{bbm}

\newcommand{\len}{\textrm{len}}

\allowdisplaybreaks

\usepackage{array}

\usepackage{cite}

\newtheorem{theorem}{Theorem}
\newtheorem{example}{Example}
\newtheorem{definition}{Definition}
\newtheorem{lemma}{Lemma}

\renewcommand{\and}{\textnormal{and}}

\newcommand{\str}{\textrm{string}}
\newcommand{\M}{\mathsf{M}}
\newcommand{\m}{\mathsf{m}}
\begin{document}

%
\title{Distributed Hypothesis Testing with Variable-Length Coding}

\author{Sadaf Salehkalaibar, \emph{IEEE Member} and Mich\`ele Wigger, \emph{IEEE Senior Member}
	\thanks{S.~Salehkalaibar is  with the Department of Electrical and Computer Engineering, College of Engineering, University of Tehran, Tehran, Iran, s.saleh@ut.ac.ir,}
	\thanks{M.~Wigger is with   LTCI,  Telecom Paris, 91120 Palaiseau, Paris, France, michele.wigger@telecom-paristech.fr.}
	\thanks{Parts of the material in this paper will be presented  at  \emph{The 2020 Workshop on Resource Allocation, Cooperation and Competition in Wireless Networks (RAWNET)}, June 2020.}
}


\maketitle

\begin{abstract} 

The problem of distributed  testing against independence  with variable-length coding is considered when the  \emph{average} and not the \emph{maximum} communication load is constrained as in previous works.  The paper characterizes the optimum type-II error exponent  of a single sensor single decision center system   given a maximum  type-I error probability when communication is  either over a noise-free rate-$R$ link or over a noisy discrete memoryless channel (DMC)  with stop-feedback. Specifically, let  $\epsilon$ denote the maximum allowed type-I error probability. Then the optimum  exponent of the system with a rate-$R$ link  under a constraint on the average communication load   coincides with the optimum  exponent  of such a system with a rate $R/(1-\epsilon)$ link under a  maximum communication load constraint. A strong converse thus does not hold under an average communication load constraint. A similar observation holds also for testing against independence over DMCs. With variable-length coding and stop-feedback and under an average communication load constraint, the optimum type-II error exponent over a DMC of capacity $C$ equals the optimum exponent under fixed-length coding and a maximum communication load constraint when communication is over a DMC of capacity $C(1-\epsilon)^{-1}$. In particular, under variable-length coding over a DMC with stop feedback a strong converse result  does not hold  and the optimum error exponent depends on the transition law of the DMC only through its capacity.  	
\end{abstract}


%
\IEEEpeerreviewmaketitle

\section{Introduction}

Consider a distributed hypothesis testing problem with a single decision center that aims at  identifying the  distribution 
governing the sources observed at the decision center itself and at various sensors. To facilitate this task, the sensors  communicate with the decision center over rate-limited links. The focus is on binary hypothesis testing problems where the sources are distributed according to one of only  two possible joint distributions, a joint  distribution $P$ under the \emph{null hypothesis} ($\mathcal{H}=H_0$) and a different joint distribution $Q$ under the \emph{alternative hypothesis} ($\mathcal{H}=H_1$). The main interest of this paper is in identifying the largest possible Stein-exponent of such systems. That is, the maximum exponential decay of the type-II error probability, i.e., the probability of deciding $H_0$ when $\mathcal{H}=H_1$, subject to a constraint on the type-I error probability, i.e., on the probability of deciding  $H_1$ when $\mathcal{H}=H_0$. Stein-exponents of distributed hypothesis testing systems have widely been studied in the information-theoretic literature, see for example  \cite{Ahlswede, Han, Amari, Wagner, TianChen, Weinberg, Lai1, Kim, Blum, Pramod, Michele, Michele3, Michele2,Piantanida, Vincent,Pierre2}. In particular, Ahlswede and  Csisz\'ar \cite{Ahlswede} have characterized the Stein-exponent of a single-sensor system where the sensor communicates with the decision center  over  a noiseless rate-limited link in the special  case of \emph{testing against independence} where  $Q$ (the joint distribution under  $H_1$) equals the product of the marginals of $P$ (the distribution under $H_0$).  The Stein exponent of this special case has also been solved in more complicated scenarios with multiple sensors \cite{Wagner},  with multiple sensors and cooperation between sensors \cite{Lai1}, with a single sensor and   successive refinement communication \cite{TianChen}, with interactive communication between sensor and decision center \cite{Kim},  with  a single sensor and multiple decision centers without and with cooperation \cite{Michele3} and \cite{Pierre2}, and in a multi-hop environment with multiple sensors and decision centers \cite{Michele}. In all these works, communication  takes place 
over rate-limited but noiseless links and the maximum allowed type-I error probability $\epsilon \to 0$. Sreekumar and G\"und\"uz \cite{Gunduz} identified the Stein exponent of the basic single-sensor single-center system when  communication takes place over a discrete memoryless channel (DMC). They showed that the Stein exponent of this setup coincides with the Stein exponent of the scenario with a noiseless link of rate equal to the capacity of the DMC. The Stein exponent thus depends on the DMC's transition law only through its capacity. The extension to multiple sensors that communicate with the single decision center over a discrete memoryless multiple-access channel was presented in \cite{SW18}. Most of  the described results can  easily be extended  also to generalized testing against independence where  the distribution $Q$ under $\mathcal{H}_1$ factorizes into the product of the marginals but not necessarily equal to the marginals of $P$ under $\mathcal{H}_1$ or to testing against conditional independence as introduced in \cite{Wagner}, see also  \cite{Michele3, Gunduz, Michele2, Yigit}. Bounds on the Stein exponents for general distributed hypothesis tests (not necessarily testing against independence or conditional independence) have also been derived for various of the described scenarios.For example, Weinberg and Kochman \cite{Weinberg} characterised the Stein-exponent under an optimal detection rule,  and Haim and Kochman recently provided  improved exponents for some general tests  with binary sources   \cite{Kochman-MAC}.

In above results, the maximum allowed type-I error probability $\epsilon$ is taken to $0$, which implies that the proofs  are  built on ``weak" converses. In contrast,  Ahlswede and Csisz\'ar showed \cite{Ahlswede} that for single-sensor single-decision center setups with a rate-limited noiseless link a ``strong" converse holds, i.e.,  the maximum type-II error exponent does not depend on $\epsilon$. This result is even more remarkable in that the optimum Stein-exponent is not known for the general hypothesis testing problem with a single noise-less link. Tian and Chen \cite{TianChen} and Cao, Zhou, and Tan \cite{Vincent} proved strong converse results  for testing against independence  in a single-sensor single-decision center setup under noiseless successive refinement communication and in a two-sensor single decision center setup with noiseless multi-hop communication. Two of the main tools for deriving strong converse results are the change of measure approach under the  \emph{$\eta$-image characterization}  \cite{Ahlswede} and  the blowing-up lemma \cite{Csiszarbook,MartonBU} or the hypercontractivity lemma \cite{Liu}.

Another line of works requires that the probability of error decays exponentially under both hypotheses and studies the pair of exponential decays that can simultaneously be achieved.  Han and Kobayashi studied the setup with one or multiple sensors that are connected over a noiseless ratelimited link with a single decision center. The extension to DMCs was proposed in \cite{Nir2}. Recently, also  finite blocklength version of this problem  was  studied in  \cite{W18}. All these works contain achievability results but no converses.

The described previous results measure  communication load in terms of the \emph{maximum} number of transmitted bits or  \emph{maximum} number of  channel uses. In this paper, we allow for variable-length coding and consider \emph{average} communication loads. When communication is over a noise-free rate-limited communication link, the average load is simply the expected number of transmitted bits, which can be different depending on the observed source sequence. When communication is over a DMC, then we allow for variable-length coding with stop-feedback from the receiver \cite{Yury} and communication load is characterised by means of expected number of channel uses. In this paper, we characterize the Stein-exponents of the single-sensor single-decision center system for testing against independence when variable-length coding is allowed and the \emph{average} communication load is constrained. The derived exponents coincide with the previously obtained exponents with fixed-length coding (and a  constraint on the maximum communication load), except that the rates/capacity of the communication links have to be multiplied by the term $(1-\epsilon)^{-1}$ where $\epsilon$ denotes the maximum allowed type-I error probability. So, variable-length coding can be seen as boosting the rate/capacity of the communication link by the factor $(1-\epsilon)^{-1}$. Notice that this implies in particular that a strong converse result does not hold under variable-length coding. Also, the maximum Stein-exponent that is achievable over a DMC depends only on the capacity of the channel but not on other properties of the DMC. 

These optimal Stein-exponents can be achieved by simple modifications of  the optimal schemes for fixed-length coding, for the latter, see for example \cite{Ahlswede, IZS2018}. The idea is to identify an event $\mathcal{S}_n$ at the sensor that happens with  probability $\epsilon'$, for $\epsilon'$ slightly smaller than the largest admissible type-I error probability  $\epsilon$. In the noiseless link setup, whenever  event $\mathcal{S}_n$ occurs, the sensor will send the single bit $0$ to the decision center, which then declares $\hat{\mathcal{H}}=H_1$. If the event $\mathcal{S}_n$ does not occur, the sensor acts as in the scheme proposed by Ahlswede and  Csisz\'ar \cite{Ahlswede}. The proposed  strategy achieves a smaller type-II error probability than the  Ahlswede-Csisz\'ar scheme and its type-I error probability   is increased at most by $\epsilon'$ (namely the probability of event $\mathcal{S}_n$) . The type-II error exponent of the modified scheme  is thus maintained and the type-I error probability  bound by $\epsilon>\epsilon'$ when the number of observations is sufficiently large. The communication rate  is decreased by a factor $(1-\epsilon)$ since no rate is required in the event $\mathcal{S}_n$. The main technical contribution in this part is the converse showing that the described simple strategy is optimal.  The converse combines  Marton's blowing up lemma \cite{MartonBU} and  a change of measure argument using the $\eta$-image characterization  similarly to \cite{Ahlswede} and \cite{TianChen}.

For the DMC, our optimal strategy takes place over two phases. In the first shorter phase, the transmitter sends a dedicated sequence $w_0^n$ if event $\mathcal{S}_n$ occurs and it sends a different sequence $w_1^n$ otherwise. The decision center performs a Neyman-Pearson test to detect which of the two sequences has been transmitted. If it detects $w_0^n$ it declares directly $\hat{\mathcal{H}}=H_1$ and sends a stop signal. Otherwise, the sensor proceeds to phase 2, where it applies the fixed-length coding scheme proposed in \cite{IZS2018} that achieves the optimal Stein-exponent under fixed-length coding  for testing against independence over a DMC. In the proposed variable-length strategy the   type-II error probability is decreased compared to the fixed-length scheme in \cite{IZS2018},  the type-I error probability is increased by at most $\epsilon'$,  and  for  large numbers  of observations,  the average number of channel uses is decreased approximately by a factor $(1-\epsilon')$. This last observation holds because   the first phase is much smaller than the second phase and transmission stops after the first phase with probability close to $\epsilon'$. We again prove the corresponding converse result. This proof requires some additional steps and considerations concerning the noisy channel law and the stop-feedback compared to the converse for the noise-less link.

\begin{figure}[t]
	\centering
	\includegraphics[scale=0.35]{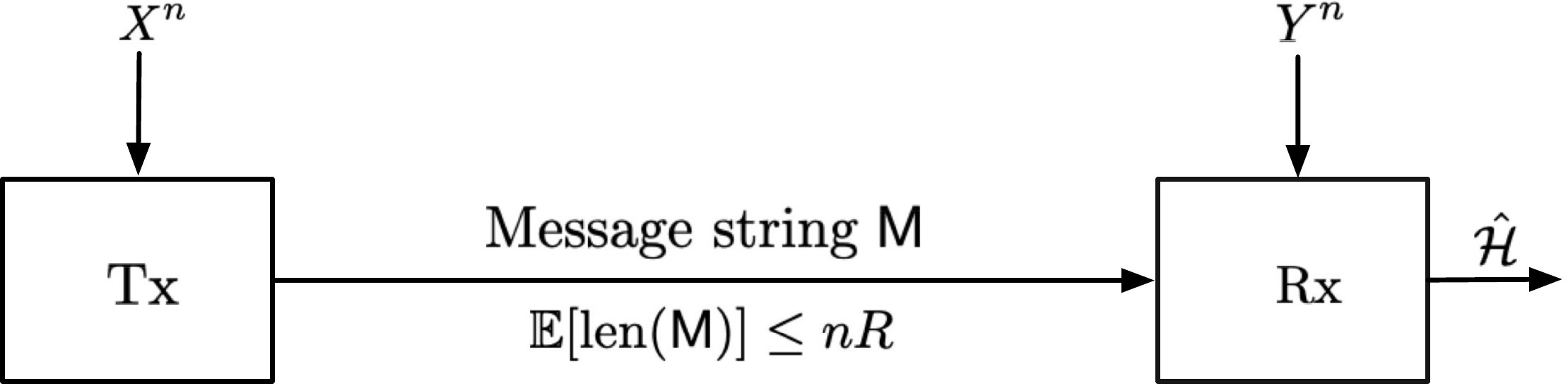}
	\caption{Variable-length hypothesis testing.}
	\label{fig1a}
\end{figure}

 The paper is organized as follows. In Section~\ref{sec:noiseless}, the distributed hypothesis testing problem over a noiseless link is studied and the result on the noisy channel is provided in Section~\ref{sec:noisy}. The proofs of the converses for the noiseless and noisy setups are provided in Sections{\tiny } \ref{sec:converse} and \ref{sec:converse2}, respectively. The paper is concluded in Section~\ref{sec:con}.
 
 We conclude the introduction with some remarks on notation. \\

\textit{Notation:}\\\\
  Random variables are denoted by capital letters, e.g., $X,$ $Y,$ and their realizations by lower-case letters, e.g., $x,$ $y$.  Script symbols  such as $\mathcal{X}$ and $\mathcal{Y}$ stand for alphabets of  random variables, and $\mathcal{X}^n$ and $\mathcal{Y}^n$ for the corresponding $n$-fold Cartesian product alphabets. We denote by $\mathcal{X}^\star$ and $\mathcal{Y}^\star$ the sets of all finite-length strings over $\mathcal{X}$ and $\mathcal{Y}$ respectively.  The set of real numbers is denoted by $\mathbb{R}$, the set of positive real numbers by $\mathbb{R}_+$, the set of integers by $\mathbb{Z}$, and the set of positive integers by $\mathbb{Z}_+$. Sequences of random variables $(X_i,...,X_j)$ and realizations $(x_i,\ldots, x_j)$ are  abbreviated by $X_i^j$ and $x_{i}^j$. When $i=1$, then we also use $X^j$ and $x^j$ instead of $X_1^j$ and $x_{1}^j$. 

We write the probability mass function (pmf) of a discrete random variable $X$  as $P_X$. The conditional pmf of $X$ given $Y$  is written as  $P_{X|Y}$. The distributions of $X^n$, $Y^n$ and $(X^n,Y^n)$  are denoted by $P_{X^n}$, $P_{Y^n}$ and $P_{X^nY^n}$, respectively. The notation $P_{XY}^n$ denotes the $n$-fold product distribution.

The term  $D(P\| Q)$ stands for  the Kullback-Leibler (KL) divergence between two pmfs $P$ and $Q$ over the same alphabet. For a given $P_X$ and a constant $\mu>0$, the set of sequences with the same type $P_X$ is denoted by $\mathcal{T}^n(P_X)$. We use  $\mathcal{T}_{\mu}^n(P_X)$ to denote the set of \emph{$\mu$-typical sequences} in $\mathcal{X}^n$:
\begin{equation}\label{eq:typical}
\mathcal{T}_{\mu}^n(P_X)=\Bigg\{x^n\colon \;\bigg| \frac{ | \{i\colon x_i=x\}|}{n}-P_X(x)\bigg|\leq \mu P_X(x), \quad\forall x\in\mathcal{X}\Bigg\},
\end{equation}
where $|\{i\colon x_i=x\}|$ is the number of positions where the sequence $x^n$ equals  $x$.
Similarly,   $\mathcal{T}_{\mu}^n(P_{XY})$ stands for the set of \emph{jointly $\mu$-typical sequences} whose definition is as in \eqref{eq:typical} with $x$ replaced by $(x,y)$. 

For any positive integer number $m\geq 1$, we use $\str(m)$ to denote the bit-string of length $\lceil \log_2(m)\rceil$ representing $m$. We further use sans serif font, e.g., $\mathsf{M}$ for a random bit-string and $\mathsf{m}$ for a deterministic bit-string, to denote finite-length bit-strings, and the function $\len(\mathsf{m})$ returns the length of a given bit-string $\mathsf{m}$. 

The Hamming distance between two sequences $x^n$ and $y^n$ is denoted by $d_{\text{H}}(x^n,y^n)$. For any $a,b\in[0, 1]$,  we denote the binary entropy function of $a$ by $h_{\text{b}}(a)$  and define $a\star b\triangleq a(1-b)+b(1-a)$.

\section{Distributed Hypothesis Testing Over a Positive-Rate Noiseless Link}\label{sec:noiseless}
\subsection{System Model}

Consider the distributed hypothesis testing problem with a transmitter and a receiver in Fig.~\ref{fig1a}. The transmitter observes the source sequence $X^n$ and the receiver observes the source sequence $Y^n$. Under the null hypothesis 
\begin{align}
	\mathcal{H}=H_0\colon \quad (X^n,Y^n)\sim \text{i.i.d.}\; P_{XY},
\end{align}
for a given pmf $P_{XY}$, whereas under the alternative hypothesis 
\begin{align}
	\mathcal{H}=H_1\colon \quad (X^n,Y^n)\sim \text{i.i.d.}\; P_{X}\cdot P_Y,
\end{align}
where $P_X$ and $P_Y$ denote the marginals of $P_{XY}$.
Upon observing $X^n$, the transmitter computes the binary message string $\M\in\{0,1\}^\star$ using a possibly stochastic encoding function \begin{align}\phi^{(n)}: \set{X}^n\to \{0,1 \}^\star,\label{phi-function}\end{align} so 
\begin{equation}
\M = \phi^{(n)}(X^n),
\end{equation}
in a way  that the expected\footnote{The expectation in \eqref{L-def} is with respect to the law of $X^n$ which equals $P_X^n$ under both hypotheses.} message length  satisfies
\begin{align}
\mathbb{E}\left[\len(\M)\right]\leq nR.\label{L-def}
\end{align}
  It then sends the binary message string $\M$ over  a noise-free bit pipe  to the receiver. 

The goal of the communication is that the receiver can determine the hypothesis $\mathcal{H}$ based on its observation $Y^n$ and its received message $\M$. Specifically, the receiver  produces the guess \begin{equation}\hat{\mathcal{H}}=g^{(n)}(Y^n,\M)
\end{equation}
using a decoding function $g^{(n)}:\mathcal{Y}^n\times \{0,1\}^\star\to \{H_0,H_1\}$.  Denoting by $\set{M}$ the set of all realizations of the binary message string $\M$, we can partition the space $\set{M}\times \set{Y}^n$ into an acceptance region for hypothesis  $H_0$
\begin{align}
	&\mathcal{A}_n \triangleq  \big\{(\m,y^n)\colon  g^{(n)}(y^n,\m)=H_0 \big\},
\end{align}
and the corresponding rejection region
 \begin{align}
 \mathcal{R}_n\triangleq (\mathcal{M}\times \mathcal{Y}^n)\backslash \mathcal{A}_n.
\end{align}

\begin{definition}\label{def} For any $\epsilon \in [0,1)$ and for a given rate $R\in \Reals_+$,  a type-II exponent $\theta\in \Reals_+$ is $(\epsilon,R)$-achievable 
	if there exists a sequence of functions $(\phi^{(n)},g^{(n)})$, such that the corresponding acceptance and rejection regions lead to a type-I  error probability 
	\begin{align}
	\alpha_{n}\eqdef \Pr[ (\M,Y^n)\in\set{R}_n | \mathcal{H}=H_0]
	\end{align} and a  type-II error probability 
	\begin{align}
	\beta_{n}\eqdef \Pr[ (\M,Y^n)\in\set{A}_n | \mathcal{H}=H_1]\end{align}
	satisfying
	\begin{align}
		\alpha_{n}&\leq \epsilon,\label{type-I-def}
	\end{align}
	and
	\begin{align}
	\liminf_{n\to\infty}\;\frac{1}{n}\log\frac{1}{\beta_{n}}\geq \theta.\label{type-II-cons}
	\end{align}
	The optimal exponent $\theta_\epsilon^*(R)$ is the supremum of all $(\epsilon,R)$-achievable type-II exponents $\theta\in \Reals_+$.
\end{definition}

\subsection{Optimal Type-II Error Exponent}

The following theorem establishes the optimal type-II error exponent $\theta_{\epsilon}^*(R)$.

\begin{theorem}\label{thm1} The optimal type-II  error exponent with variable-length coding  is
	\begin{align}
	\theta^*_{\epsilon}(R) = \max_{\substack{P_{U|X}\colon \\R\geq (1-\epsilon)I(U;X)}}I(U;Y).
	\end{align}
\end{theorem}
\begin{IEEEproof} Here we only prove  achievability. The converse is more technical and proved in Section~\ref{sec:converse}.   
	
	\textbf{Achievability}: Fix a large blocklength $n$, a small number $\mu\in(0, \epsilon)$, and a conditional pmf $P_{U|X}$ such that:
	\begin{align}\label{eq:rateR}
	R= (1-\epsilon+\mu)I(U;X)+\mu.
	\end{align}
	 Then define  the joint pmf
	\begin{align}
	P_{UXY}\triangleq P_{U|X}\cdot P_{XY}
	\end{align}
	and randomly generate an $n$-length  codebook $\mathcal{C}_U$ of rate $R$  
	 by picking all entries i.i.d. according to the marginal pmf $P_U$. The realization of the codebook 
	 		\begin{align}
	 		\mathcal{C}_U\triangleq \left\{ u^n(m)\colon m\in\left\{1,\ldots,\lfloor 2^{nR}\rfloor\right\}  \right\}
	 		\end{align} is revealed to all terminals.

	Finally, choose a subset $\mathcal{S}_n\subseteq \mathcal{T}_{\mu/2}^{(n)}(P_X)$ such that \begin{equation}
	\Pr\left[X^n\in \mathcal{S}_n\right]=\epsilon-\mu.
	\end{equation}

	
	\underline{\textit{Transmitter}}: Assume it observes $X^n=x^n$. If 
	\begin{equation}
	x^n \notin \mathcal{S}_n,
	\end{equation}
	 it looks for an index $m\in \{1,\ldots,\lfloor 2^{nR}\rfloor\}$ such that 
\begin{equation}(u^n(m),x^n)\in\mathcal{T}_{\mu/2}^n(P_{UX}). 
\end{equation}If  successful, it picks  one of these indices  uniformly at random and sends  the binary representation of the chosen index  over the noiseless link. So, if the chosen index is $m^*\in \{1,\ldots,\lfloor 2^{nR}\rfloor\}$, it sends the corresponding length-$nR$ bit-string
\begin{equation}
\M = \str(m^*).
\end{equation}
Otherwise it sends the single bit $\M=[0]$. 
	
	\underline{\textit{Receiver}}: If it  receives  the single bit $\M=[0]$, it declares $\hat{\mathcal{H}}=H_1$. Otherwise, if the bit string $\M$ corresponds to a given index $m\in \{1,\ldots,\lfloor 2^{nR}\rfloor\}$,  it checks whether $(u^n(m),y^n)\in\mathcal{T}_{\mu}^n(P_{UY})$. If successful, it declares $\hat{\mathcal{H}}=H_0$, and otherwise it declares $\hat{\mathcal{H}}=H_1$.
	
	\underline{\textit{Analysis}}: The proposed coding scheme is analyzed when averaged over the random codeconstruction. By standard arguments it can then be concluded that the desired exponent is achievable also for at least one realizations of the codebooks.
	
	Since a single bit is sent when $x^n \in \mathcal{S}_n$, the expected message length can be bounded  as:
	\begin{IEEEeqnarray}{rCl}
		\mathbb{E}\left[\len(\M)\right] &= & \Pr[X^n\in \mathcal{S}_n]\cdot \mathbb{E}\left[\len(\M)|X^n\in \mathcal{S}_n\right]+\Pr[X^n\notin \mathcal{S}_n]\cdot \mathbb{E}\left[\len(\M)|X^n\notin \mathcal{S}_n\right]\\
		&\leq& (\epsilon-\mu)\cdot 1 + (1-\epsilon+\mu)\cdot n(I(U;X)+\mu) ,
	\end{IEEEeqnarray}
which for sufficiently large $n$ is further bounded as (see  \eqref{eq:rateR}):
\begin{equation}\label{eq:LR}
		\mathbb{E}\left[\len(M)\right] < nR.
\end{equation}

To bound the type-I and type-II error probabilities, we notice that when $x^n \notin \mathcal{S}_n$, the scheme  coincides with the one proposed by Ahlswede and Csisz\`ar in \cite{Ahlswede}. When $x^n \in \mathcal{S}_n$, the transmitter sends the single bit $\M=[0]$ and the receiver declares $H_1$. The type-II error probability of our scheme is thus no larger than the type-II error probability of the Ahlswede-Csisz\`ar scheme in \cite{Ahlswede}, and the type-I error probability is at most 
$\Pr[X^n\in \mathcal{S}_n]=\epsilon-\mu$ larger than for this Ahlswede-Csisz\`ar scheme. 
Since the type-I error probability of  the Ahlswede-Csisz\`ar scheme tends to 0 as $n \to \infty$ \cite{Ahlswede},
 the type-I error probability here is bounded by
 $\epsilon$, for sufficiently large values of $n$ and all choices of $\mu \in(0,\epsilon)$.
 Combining these considerations  with  \eqref{eq:LR}, and  letting $n\to\infty$ and $\mu \to 0$  establishes the achievability part of the proof.
\end{IEEEproof}

\medskip
For comparison, recall the result in \cite{Ahlswede} which showed that under fixed-length coding, i.e., when instead of the  average message length constraint \eqref{L-def} only the  maximum message length is constrained by $nR$,  the optimal type-II error exponent equals:
\begin{equation}
\theta^*_{\textnormal{FL}}(R) = \max_{\substack{P_{U|X}\colon \\R\geq I(U;X)}}I(U;Y).
\end{equation}
Under fixed-length coding, the optimal type-II error exponent  does hence  not depend on the maximum allowed type-I error probability $\epsilon$ and we say that a ``strong converse" holds. Our result shows that such a ``strong converse" does not hold under variable-length coding and also quantifies the gain in type-II error exponent as a function of the maximum allowed type-I error probability. The gain of variable-length coding is also illustrated at hand of two concrete examples.
\medskip

{\begin{example}\label{ex} Suppose that the source alphabets are  binary with $P_X=P_Y\sim \text{Bern}(\frac{1}{2})$ and the conditional pmf $P_{Y|X}$ is given by
		\begin{IEEEeqnarray}{rCl}
			P_{Y|X}(y|x)&=&\begin{pmatrix} 1-\alpha & \alpha\\ \alpha & 1-\alpha\end{pmatrix},
		\end{IEEEeqnarray}
		where $0\leq \alpha<\frac{1}{2}$. We can write the following set of inequalities:
		\begin{IEEEeqnarray}{rCl}
			\theta^*_{\epsilon}(R)
			&=&\max_{\substack{P_{U|X}:\\R\geq (1-\epsilon) I(U;X)}}I(U;Y)\\&=&\max_{\substack{P_{U|X}:\\1-h_{\text{b}}(X|U)\leq \frac{R}{1-\epsilon}}}	1-h_{\text{b}}(Y|U)\nonumber\\\\
			&= & 1-h_{\text{b}}\left(h_{\text{b}}^{-1}\left(1-\frac{R}{1-\epsilon}\right)\star \alpha\right)
		\end{IEEEeqnarray}
		Notice that the last equality follows from Ms. Gerber's lemma \cite[p. 19]{ElGamal}.
		
		Following similar steps, it can be shown that the optimal type-II error exponent under variable-length coding evaluates to
		\begin{IEEEeqnarray}{rCl}
			\theta^{*}_{\text{FL}}(R) = 1-h_{\text{b}}\left(h_{\text{b}}^{-1}\left(1-R\right)\star \alpha\right).
		\end{IEEEeqnarray}
		Fig.~\ref{fig2} shows the optimal error exponents $\theta^*_{\epsilon}(R)$ and $\theta^*_{\text{FL}}(R)$ in functions of the parameter $\alpha$ for $\epsilon=0.1$  and $R=0.8$. The gain of variable-length coding compared to fixed-length coding seems to be particularly pronounced for small values of $\alpha$, where the sources are highly correlated under the null hypothesis $H_0$.  
		\begin{figure}[t]
			\centering
			\includegraphics[scale=0.35]{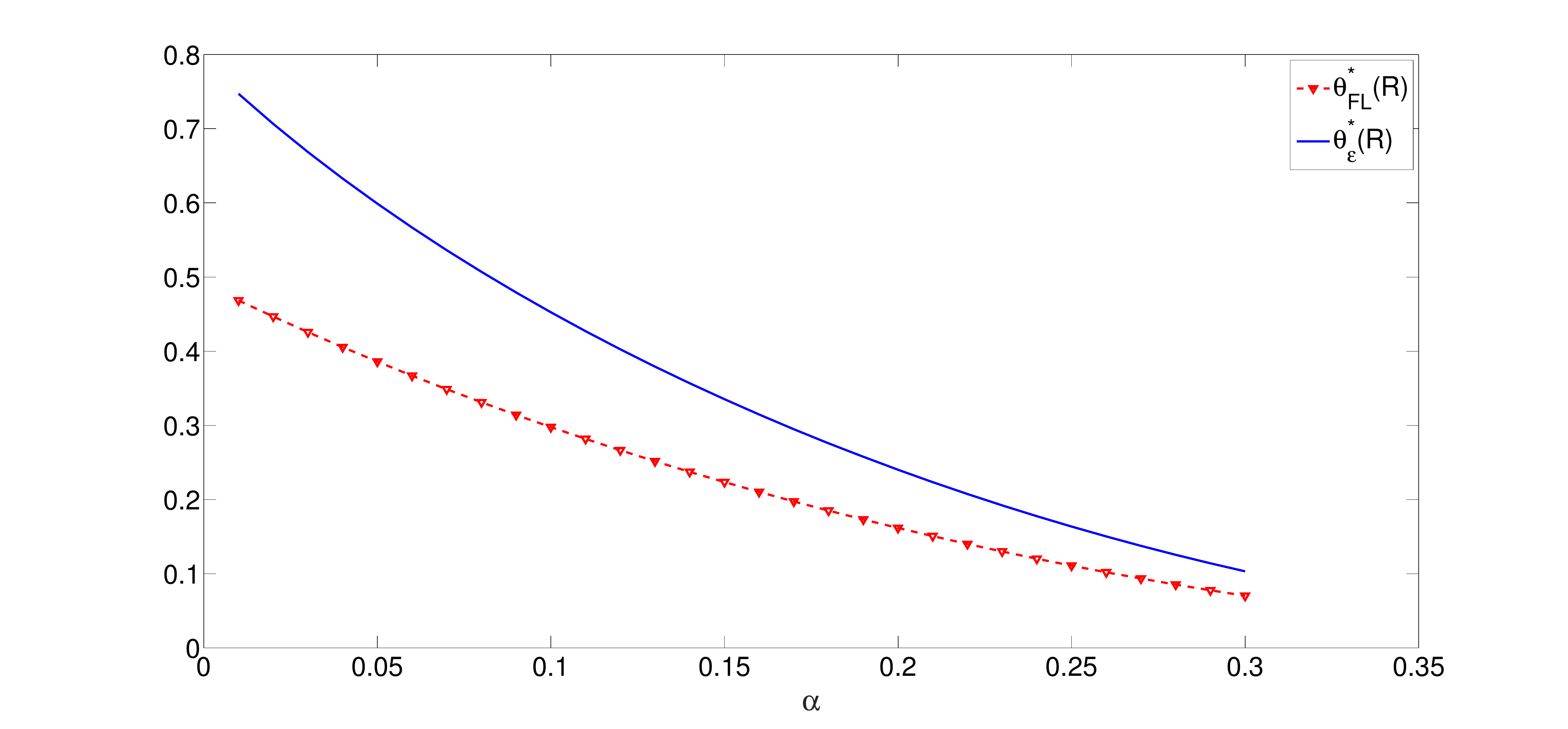}
			\caption{Comparison of fixed-length and variable-length codings for Example~\ref{ex}.}
			\label{fig2}
		\end{figure}
		
		%
\end{example} 
\medskip
\begin{example}\label{ex2} Given  $\rho \in [0,1]$, define the two covariance matrices
	\begin{align}
	\mathbf{K}_{XY}^0=\left[ \begin{array}{cc} 1 & \rho \\ \rho & 1\end{array} \right] \qquad \textnormal{and}\qquad 	\mathbf{K}_{XY}^1=\left[ \begin{array}{cc} 1 & 0 \\ 0  & 1 \end{array} \right].
	\end{align}
	
	Under the null hypothesis, 
	\begin{align}
	\mathcal{H}=H_0\colon\qquad (X,Y)\sim \mathcal{N}(0,\mathbf{K}_{XY}^0),
	\end{align}
	and under the alternative hypothesis, 
	\begin{align}
	\mathcal{H}=H_1\colon\qquad(X,Y)\sim \mathcal{N}(0,\mathbf{K}_{XY}^1).
	\end{align}

The above setup can model a communication scenario with a jammer. Under the null hypothesis, the jammer interferes with the communication and the observations at the transmitter and receiver are correlated with each other where the correlation is modelled by the parameter $\rho$. Under the alternative hypothesis, the jammer remains silent and the observations $X^n$ and $Y^n$  are independent of each other. The goal of the system is to detect the presence of the jammer.

To characterize the optimal type-II error exponent in the above example, notice that under $\mathcal{H}=H_0$, one can write $Y=\rho X +Z$ with $Z$  a zero-mean Gaussian random variable of variance $1-\rho^2$ and independent of  $X$. Consider the following set of inequalities: 	
\begin{IEEEeqnarray}{rCl}
	\theta^*_{\epsilon}(R)
	&=&\max_{\substack{P_{U|X}:\\R\geq (1-\epsilon) I(U;X)}}I(U;Y)\\
	&=& \frac{1}{2}\log (2\pi e)-\min_{\frac{R}{1-\epsilon}\geq  \frac{1}{2}\log (2\pi e)-h(X|U)}h(Y|U)\\
	&\leq & \frac{1}{2}\log (2\pi e)-\min_{\frac{R}{1-\epsilon}\geq  \frac{1}{2}\log (2\pi e)-h(X|U)}\frac{1}{2} \log \left( 2 \pi e \left(\frac{1}{2\pi e} 2^{2h(\rho X|U)} + (1- \rho^2)\right) \right)\\
	&=& \frac{1}{2}\log (2\pi e)-\min_{\frac{R}{1-\epsilon}\geq  \frac{1}{2}\log (2\pi e)-h(X|U)}\frac{1}{2} \log \left( 2 \pi e \left(\frac{\rho^2}{2\pi e} 2^{2h( X|U)} + (1- \rho^2)\right)\right)\\
	&=& \frac{1}{2}\log \left(\frac{1}{1-\rho^2+\rho^2\cdot 2^{-\frac{2R}{1-\epsilon}}}\right),
	\end{IEEEeqnarray}
where the inequality follows from the entropy-power inequality (EPI) \cite[pp. 22]{ElGamal}. Notice that the above exponent can be achieved by choosing $U$ jointly Gaussian with $X$. 

Following similar steps, one can show that the optimal type-II error exponent under variable-length coding evaluates to:
\begin{IEEEeqnarray}{rCl}
	\theta^{*}_{\text{G,FL}}(R)=\frac{1}{2}\log \left(\frac{1}{1-\rho^2+\rho^2\cdot 2^{-2R}}\right).
	\end{IEEEeqnarray}
Fig.~\ref{fig2b} shows the optimal error exponents $\theta^*_{\epsilon}(R)$ and $\theta^*_{\text{G,FL}}(R)$ versus parameter $\rho$ for $\epsilon=0.1$  and $R=0.8$. For large values of the parameter $\rho$ where the sources are highly correlated under the null hypothesis, variable-length coding outperforms fixed-length coding.
	\end{example}
\begin{figure}[t]
	\centering
	\includegraphics[scale=0.35]{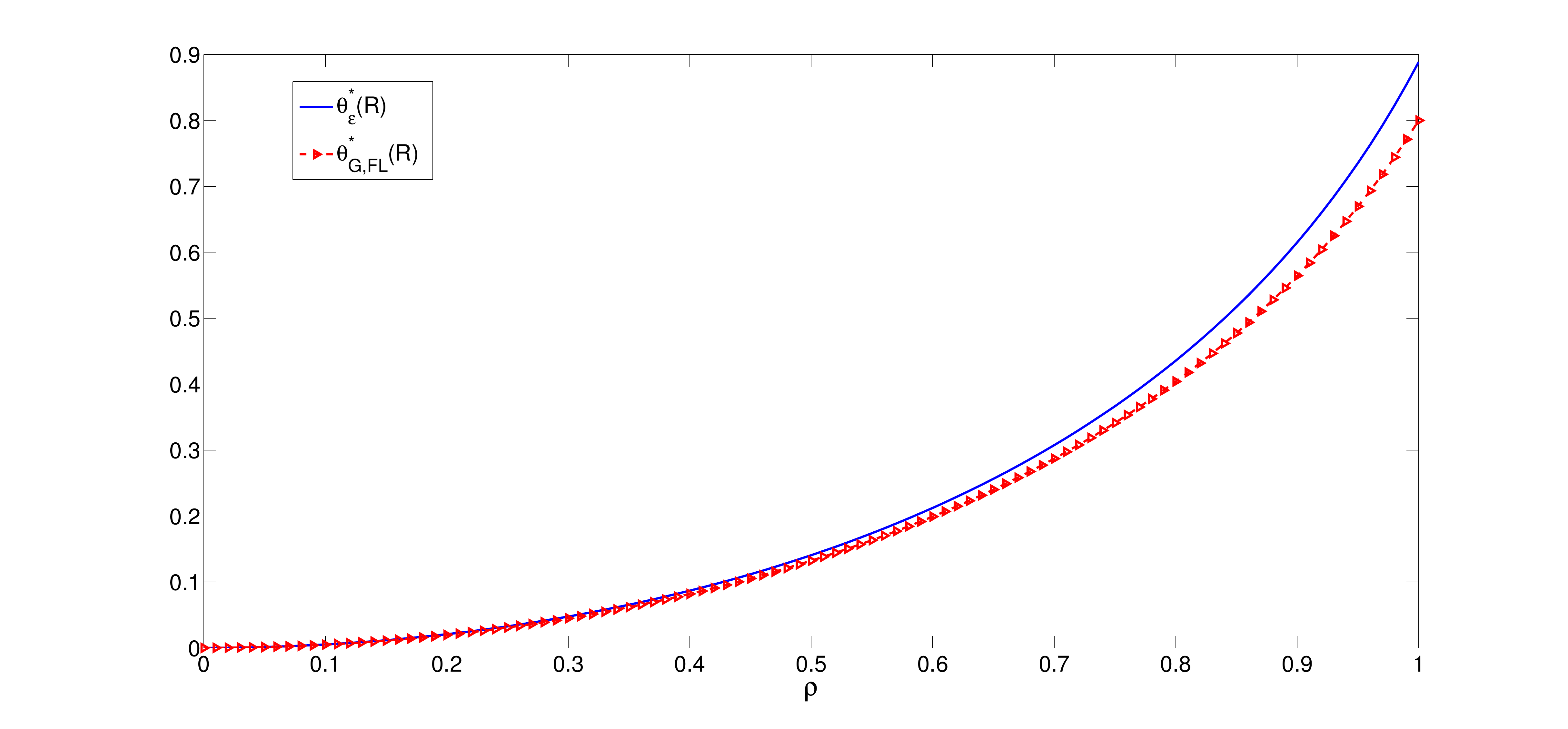}
	\caption{Comparison of fixed-length and variable-length codings for Example~\ref{ex2}.}
	\label{fig2b}
\end{figure}

\section{Testing Over a Discrete Memoryless Channel (DMC)}\label{sec:noisy}
\subsection{System Model}
%

\begin{figure}[b]
	\centering
	\includegraphics[scale=0.35]{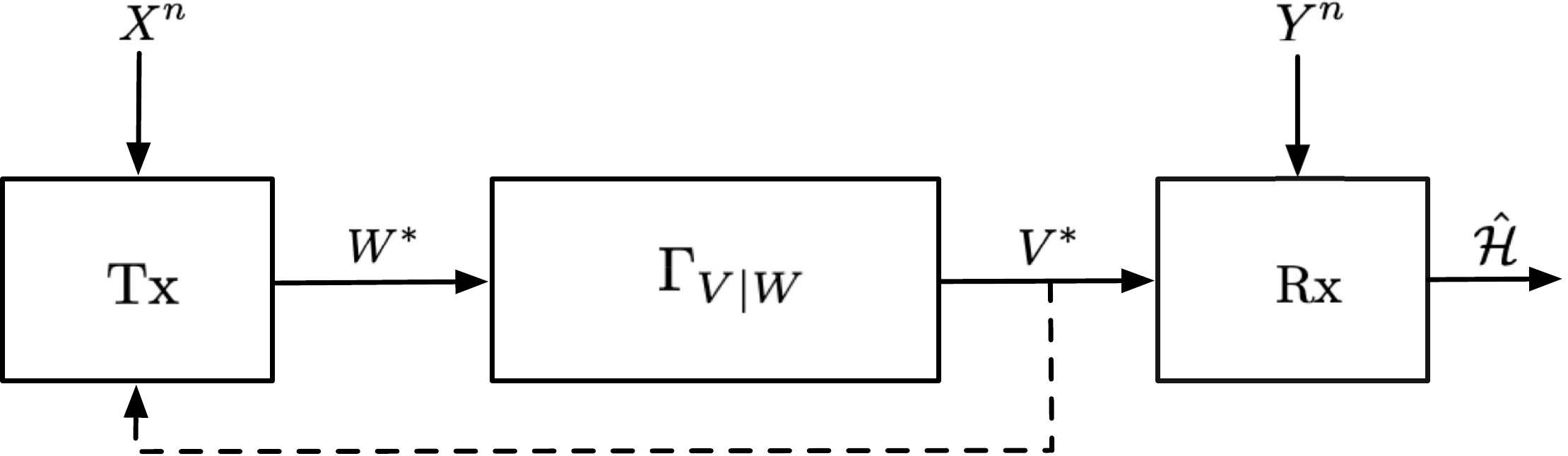}
	\caption{Hypothesis testing over a noisy channel with variable-length coding and stop feedback.}
	\label{fig1}
\end{figure}

Consider a hypothesis testing system with a single transmitter and a single receiver where communication is over a discrete memoryless channel (DMC) with input alphabet $\mathcal{W}$, output alphabet $\mathcal{V}$, and transition law $\Gamma_{V|W}(\cdot|\cdot)$. The number of channel uses is a random quantity,  because the transmitter stops transmission only after receiving a feedback signal  from the receiver. This stop feedback-signal is without error or delay.

As in the previous section, the transmitter observes the source sequence $X^n$ and the receiver observes the side-information sequence $Y^n$, where 
\begin{align}
\textnormal{under }\;	\mathcal{H}=H_0\colon \quad (X^n,Y^n)\sim \text{i.i.d.}\; P_{XY},
\end{align}
and
\begin{align}
\textnormal{under }\;	\mathcal{H}=H_1\colon \quad (X^n,Y^n)\sim \text{i.i.d.}\; P_{X}\cdot P_Y.
\end{align}

Based on the source sequence $X^n$, the transmitter  generates an infinite-length stream 
	\begin{equation}W'^\infty(X^n)=W_1', W_2',\ldots
	\end{equation}and for each channel use prior to the stop-feedback, it sends the corresponding symbol of the sequence $W'^\infty(X^n)$ over the channel.    For each time-instant $k$, let  $L_k=1$ indicate that the receiver has not yet sent the stop-symbol, and  $L_k=0$ otherwise. We then have for the  time-$k$ channel input $W_k$:
	\begin{equation}
	W_k = W_k' \qquad \textnormal{ if } L_k=1, \textnormal{ for } k=1,2,\ldots
	\end{equation}
	and the  transmission duration is
		\begin{equation}\label{eq:trans_duration}
	\tau_n \triangleq \min \{ k \colon L_k =0\}.
	\end{equation}
	
	The receiver observes the random channel outputs $V_1,V_2, \ldots, V_{\tau_n}$ corresponding to the inputs $W_1,W_2,\ldots, W_{\tau_n}$ fed to the given DMC $\Gamma_{V|W}$. 
	At each time $k=1,2,\ldots$, the receiver decides whether the communication should continue ($L_k=1$)  or not ($L_k=0$). For simplicity, we assume that the  decision  $L_k$ is only a function of the first $k-1$ channel outputs $V_1,\ldots, V_{k-1}$ but not of $Y^n$. This models for example  a situation where the receiver learns the side-information  $Y^n$ only after the communication has terminated. Thus, 	in our scenario:\begin{equation}
	L_k = e_k^{(n)}(V^k), 
	\end{equation}
	for each $k=1,2, \ldots$ and some stopping function $e_k\colon \mathcal{V}^k\to \{0,1\}$.  
The stopping functions   determine  the set of all output strings  for which the receiver stops the transmission: 
	\begin{equation}
	\mathcal{V}_{\textnormal{stop}}  \triangleq \left\{ v^\tau \in \mathcal{V}^\star \colon e_{\tau}^{(n)}(v^{\tau})=0 \; \textnormal{and} \; e^{(n)}_{\tau-1}(v^{\tau-1})=1\right\}.
	\end{equation}
	where here $v^{\tau-1}$ denotes the first $\tau-1$ symbols of $v^{\tau}$.

	Once transmission stops, the receiver has  observed the channel outputs $V^{\tau_n}\in \mathcal{V}_{\textnormal{stop}}$ and the  side-information $Y^n$. Based on these observations it has to  guess  the hypothesis $\hat{\mathcal{H}}=H_0$ or  $\hat{\mathcal{H}}=H_1$. To this end, it chooses a subset $\mathcal{A}_n \subset \mathcal{V}_{\textnormal{stop}} \times \mathcal{Y}^n$, which we call the acceptance region, and  it decides on  $\hat{\mathcal{H}}=H_0$  whenever $(V^{\tau_n},Y^n)\in\mathcal{A}_n$. Conversely,  it decides on  $\hat{\mathcal{H}}=H_1$  whenever $(V^{\tau_n},Y^n)$ lies in the complement $\mathcal{R}_n \triangleq (\mathcal{V}_{\textnormal{stop}} \times \mathcal{Y}^n) \backslash \mathcal{A}_n$, which we call the rejection region.

The type-I error probability is then defined as:
\begin{align}
\alpha_{n}\eqdef 
P_{V^{\tau_n}Y^n}(\set{R}_{n})=1-  P_{V^{\tau_n}Y^n}(\set{A}_{n}),
\end{align} 
and the type-II error probability as:
\begin{align}
\beta_{n}\eqdef 
P_{V^{\tau_n}} P_{Y}^n(\set{A}_{n}).
\end{align}

\begin{definition}\label{def} For any $\epsilon \in [0,1)$ and a given bandwidth mismatch factor $\kappa \in \Reals_+$, we say that a type-II  error exponent $\theta\in \Reals_+$ is $(\epsilon,\kappa)$-achievable 
	if there exists a sequence of encoding functions, stopping functions and acceptance regions $\big\{\{\Phi_k^{(n)}\}_{k\geq 1},\{e_k^{(n)}\}_{k\geq 1},\mathcal{A}_n\big\}_{n\geq 1}$, such that the corresponding sequences of type-I  and type-II  error probabilities 
	satisfy 
	\begin{align}\label{eq:t1}
	\alpha_{n}&\leq \epsilon,
	\\
	\liminf_{n\to\infty}\;\frac{1}{n}\log\frac{1}{\beta_{n}} &\geq \theta,
	\end{align}
	and the average transmission duration $\mathbb{E}[\tau_n]$  satisfies 
	\begin{align}
	\limsup_{n\to\infty}\frac{\mathbb{E}\left[ \tau_n \right]}{n}\leq \kappa.\label{stop-time}
	\end{align}
	
	Given  $\kappa\in \Reals_+$, the optimal exponent $\theta_{\textnormal{DMC},\epsilon}^*(\kappa)$ is the supremum of all $(\epsilon,\kappa)$-achievable type-II error exponents $\theta\in \Reals_+$.
\end{definition}

\subsection{Optimal Error Exponent}
\begin{theorem} \label{thm:noisy}The optimal type-II exponent over a DMC $(\mathcal{W}, \mathcal{V}, \Gamma_{V|W})$  with variable-length coding and  stop feedback is:
	\begin{align}\label{eq:optimalDMC}
	\theta^*_{\textnormal{DMC},\epsilon}(\kappa)= \max_{\substack{P_{U|X}:\\\kappa C\geq (1-\epsilon) I(U;X)}}I(U;Y),
	\end{align}
	where $C$ denotes the capacity of the DMC $(\mathcal{W}, \mathcal{V}, \Gamma_{V|W})$.
\end{theorem}
\begin{IEEEproof}  The converse is proved 
	in Section~\ref{sec:converse2}. The achievability in the following subsection~\ref{sec:achDMC}.

Under fixed-length coding, the optimal type-II error exponent was  derived in the asymptotic regime $\epsilon \to 0$ \cite{Gunduz}:

\begin{align}\label{eq:optimalDMC}
	\theta^*_{\textnormal{DMC,FL}}(\kappa):= \max_{\substack{P_{U|X}:\\\kappa C\geq  I(U;X)}}I(U;Y).
	\end{align}
It can be shown that the same exponent is optimal for arbitrary $\epsilon \in (0,1)$ and thus a ``strong converse" holds under fixed-length coding. In contrast, our result shows that under variable-length coding a ``strong converse" does not hold and it characterises the gain in optimal type-II error exponent when a type-I error probability of $\epsilon>0$ is tolerated.

\subsection{Coding Scheme Achieving the Optimal Exponent}\label{sec:achDMC}

We now prove achievability of the exponent in \eqref{eq:optimalDMC}.
 Choose two  different symbols $w_0,w_1\in\mathcal{W}$ such that the KL-divergence of the output distributions induced by these inputs is positive, i.e., such that 
 \begin{equation}
 D( \Gamma_{w_0}\|  \Gamma_{w_1})>0,
 \end{equation}
 where
 \begin{align}
 \Gamma_{w_0}(\cdot)\triangleq\Gamma(\cdot|w_0),\qquad \Gamma_{w_1}(\cdot)\triangleq\Gamma(\cdot |w_1).
 \end{align}
Further, choose a positive number $\epsilon'\in(0,\epsilon)$ close to $\epsilon$
and a function $q\colon \mathbb{Z}^+ \to \mathbb{Z}^+$ that satisfies the following two conditions:
\begin{IEEEeqnarray}{rCl}\label{eq:limit_q}
\lim_{n\to \infty} q(n) &= &\infty\\
\lim_{n\to \infty} \frac{q(n) }{n} &= &0.
\end{IEEEeqnarray}
Define
\begin{equation}
\mu\triangleq \epsilon-\epsilon'.
\end{equation}

Fix two pmfs $P_{U|X}$ and $P_W$ and a positive rate $R$ so that the following two conditions hold:
\begin{align}
R&=I(U;X)+\mu,\label{R-define}\\
R&<\frac{\kappa}{1-\epsilon'}I(W;V).
\end{align}
Define $P_{UX}\triangleq P_{U|X}\cdot P_X$ and $P_{WV}\triangleq P_W\cdot \Gamma_{V|W}$.
 
Fix now a large blocklength $n$  and generate two codebooks 
\begin{IEEEeqnarray}{rCl}
	\mathcal{C}_U&\triangleq &\left\{ u^{n}(m)\colon m\in\{1,\ldots,\lfloor 2^{nR}\rfloor \} \right\},\\
	\mathcal{C}_W&\triangleq & \left\{ {w}^{n'}(m)\colon m\in\{0,\ldots,\lfloor 2^{nR}\rfloor\} \right\},
\end{IEEEeqnarray}
where 
\begin{equation}\label{eq:nprime}
n'\triangleq \frac{n\kappa}{1-\epsilon'},  
\end{equation}
and  where the entries of the two codebooks are picked i.i.d. according to the pmfs $P_U$ and $P_W$, respectively. Furthermore, choose a subset $\mathcal{S}_n\subseteq \mathcal{T}_{\mu/2}^{(n)}(P_X)$ such that \begin{equation}
 \Pr\left[X^n\in \mathcal{S}_n\right]=\epsilon'.
 \end{equation}

The coding scheme decomposes into two phases. \\
\noindent\underline{\textbf{Phase 1}}: Consists of the first  $q(n)$ channel uses. 

\underline{\textit{Transmitter}}: Given that it observes $X^n=x^n$, the transmitter sends the $q(n)$ inputs
\begin{equation}
(W_1, \ldots, W_{q(n)}) = \begin{cases} w_1^{\otimes q(n)}, & \textnormal{ if } X^n\in \mathcal{S}_n,
\\w_0^{\otimes q(n)}, & \textnormal{ otherwise},
\end{cases}
\end{equation}
where for any input symbol $w\in \mathcal{W}$,
\begin{align}w^{\otimes j}\triangleq (\underbrace{w,\ldots,w}_{j\;\text{times}}), \quad j \in \mathbb{Z}^+.\end{align}

\underline{\textit{Receiver}}: Upon observing the first $q(n)$ channel outputs $V_1, \ldots, V_{q(n)}$,  the receiver performs a Neyman-Pearson test to decide on whether the transmitter sent $w_0^{\otimes q(n)}$ or $w_1^{\otimes q(n)}$. This test only depends on the channel outputs but not on the receiver's side-information $Y^n$. The threshold of the  test is set so that the probability of declaring $w_1^{\otimes q(n)}$ when $w_0^{\otimes q(n)}$ was sent, equals $\mu/3$.

If the receiver detects $w_1^{\otimes q(n)}$, then it decides on 
\begin{equation}
\hat{\mathcal{H}}=H_1
\end{equation}
and sends the stop feedback $L_{q(n)}=0$. to the transmitter which stops 
 transmission.

If the receiver instead detects  $w_0^{\otimes q(n)}$,  then it waits to make a decision and also does not send the stop feedback. Both the  transmitter and the receiver move on to Phase~2. In this second phase, the receiver  will ignore all outputs from the first phase.

%
%

\underline{\textbf{Phase 2}}: 
This second phase consists of $n'$ channel uses.

\underline{\textit{Tansmitter}}: It looks for a codeword $u^n(m)$ such that  $(u^n(m),x^n)\in\mathcal{T}_{\mu/2}^n(P_{UX})$. If no such index exists, it sends $w^{n'}(0)$ over the channel. If one or multiple such indices can be found, the transmitter picks $m^*$ uniformly at random among them and sends the corresponding channel codeword $w^{n'}(m^*)$ over the channel. 

\underline{\textit{Receiver}}: Let   $v_{2}^{n'}$ denote the $n'$ channel outputs observed at the receiver during this second phase.  The receiver looks for a  unique index $m\in\{0,\ldots, \lfloor 2^{nR}\rfloor\}$ such that  
\begin{equation}(w^{n'}(m),v^{n'})\in\mathcal{T}_{\mu}^{n'}(P_{WV}) .
\end{equation}
If $m=0$ or none of the indices satisfy the condition, the receiver declares $\hat{\mathcal{H}}=H_1$. Otherwise, it    produces the decoded message $\hat{M}\in\{1,\ldots, \lfloor 2^{nR}\rfloor\}$ equal to the unique index $m$ and proceeds with the hypothesis test: if $\hat{M}=\hat{m}$ and 
\begin{equation} 
(s^n(\hat{m}),y^n)\in\mathcal{T}_{\mu}^n(P_{SY}),
\end{equation}
then the receiver declares $\hat{\mathcal{H}}=\mathcal{H}_0$, otherwise it declares $\hat{\mathcal{H}}=\mathcal{H}_1$.

 In any case it sends the stop-feedback to stop the transmission, $L_{q(n)+n}=0$.

\noindent \underline{\textit{Analysis:}}
We first analyze the 
expected transmission duration. Notice that for the described scheme, the transmission duration does not depend on the hypothesis, because it only depends on $X^n$ and the DMC which have same distributions under both hypotheses. 


When transmission goes to phase $2$, i.e., $L_{q(n)}=1$, then the transmission duration equals $\tau_n=n'+q(n)$ and when $L_{q(n)}=0$, then $\tau_n=q(n)$. Therefore, 
\begin{IEEEeqnarray}{rCl}
	\mathbb{E}\left[ \tau_n \right]
	&=& q(n)+n'\cdot \Pr\left[L_{q(n)}=1\right].\label{t-step1}
	\end{IEEEeqnarray}
	To bound $\Pr\left[L_{q(n)}=1\right]$, we notice that 
  by the way we set the threshold for the Neyman-Pearson test: \begin{equation}\Pr\left[L_{q(n)}=1\Big|(W_1,\ldots, W_{q(n)})=w_0^{\otimes q(n)}\right]=1-\mu/3.
	\end{equation}
	Moreover, by the property of the Neyman-Pearson test, when $n$ (and thus $q(n)$) is sufficiently large,  the probability of going to phase $2$ after sending $w_1^{\otimes q(n)}$ in phase 1 is bounded as:
	\begin{IEEEeqnarray}{rCl}
	2^{-q(n)\left(D(\Gamma_{w_0}\|\Gamma_{w_1})+\mu\right)}	\leq \Pr\left[L_{q(n)}=1\Big|(W_1,\ldots, W_{q(n)})=w_1^{\otimes q(n)}\right] \leq 2^{-q(n)\left(D(\Gamma_{w_0}\|\Gamma_{w_1})-\mu\right)}.
		\end{IEEEeqnarray}
Using that  in phase 1 the sequence $w_1^{\otimes q(n)}$ is sent with probability $\epsilon'$ and the sequence $w_0^{\otimes q(n)}$ with probability $1-\epsilon'$, we conclude that
\begin{IEEEeqnarray}{rCl}
	 \Pr\left[L_{q(n)}=1\right] &=& \Pr\left[(W_1,\ldots, W_{q(n)})=w_0^{\otimes q(n)}\right]\cdot \Pr\left[L_{q(n)}=1\Big|(W_1,\ldots, W_{q(n)})=w_0^{\otimes q(n)}\right] \nonumber \\
	  && +\Pr\left[(W_1,\ldots, W_{q(n)})=w_1^{\otimes q(n)}\right]\cdot \Pr\left[L_{q(n)}=1\Big|(W_1,\ldots, W_{q(n)})=w_1^{\otimes q(n)}\right]\\
	 &\leq & (1-\epsilon')\cdot  (1-\mu/3 )+ \epsilon' \cdot 2^{-q(n)( D( \Gamma_{w_0} \|\Gamma_{w_1}) -\mu) }.\label{t-step2}
	\end{IEEEeqnarray}
Since $q(n) \to \infty$ as $n\to \infty$, for sufficiently large $n$:
\begin{IEEEeqnarray}{rCl}
	\Pr\left[L_{q(n)}=1\right] \leq 1-\epsilon',\end{IEEEeqnarray}
and by
\eqref{t-step1}: 
\begin{IEEEeqnarray}{rCl}
	\mathbb{E}\left[\tau_n\right]&\leq q(n)+ (1-\epsilon') n'.
	\end{IEEEeqnarray}
Dividing both sides of the above inequality by $n$ and letting $n\to\infty$, we obtain:
\begin{IEEEeqnarray}{rCl}
	\lim_{n\to\infty} \frac{\mathbb{E}\left[\tau_n\right]}{n}\leq \kappa.
	\end{IEEEeqnarray}

\underline{\textit{Analysis of error probabilities}}: The analysis is performed averaged over the random codebooks.  To  simplify notation, we introduce a virtual transmitter/receiver pair that always continues to Phase 2 (irrespective of the outcome of the Neyman-Pearson test), and we denote by $\hat{M}_2$  the decoded message produced by this virtual receiver  by $\hat{H}_2$ its guess at the end of  Phase 2. Notice that when $L_{q(n)}=1$, then $\hat{\mathcal H}_{2}=\hat{\mathcal H}$.

Consider first the type-I error probability. When $L_{q(n)}=0$ then $\hat{\mathcal{H}}=H_1$ with probability 1. Therefore,  for sufficiently large values of $n$:
\begin{IEEEeqnarray}{rCl}
	\lefteqn{
	\Pr\left[\hat{\mathcal{H}}=H_1 \Big| \mathcal{H}=H_0\right] }\quad \\
&=& \Pr\left[ L_{q(n)}=0 \Big| \mathcal{H}=H_0 \right] + \Pr\left[ \hat{\mathcal{H}}=H_1 , L_{q(n)}=1\Big| \mathcal{H}=H_0 \right]\\
%
 	&=&
	\Pr\left[ L_{q(n)}=0, (W_1,\ldots, W_{q(n)}) = w_0^{\otimes q(n)} 	\Big| \mathcal{H}=H_0\right]
 	 +\Pr\left[L_{q(n)}=0, (W_1,\ldots, 
 	W_{q(n)}) = w_1^{\otimes q(n)}\Big| \mathcal{H}=H_0\right] 
 	\nonumber \\ 
 	&&	+		\Pr\left[ \hat{\mathcal{H}}=H_1, L_{q(n)}=1 \Big| \mathcal{H}=H_0\right]\\
		&\leq&
	\Pr\left[	 L_{q(n)}=0\Big|(W_1,\ldots, W_{q(n)}) = w_0^{\otimes q(n)}, \mathcal{H}=H_0\right] +\Pr\left[(W_1,\ldots, W_{q(n)}) = w_1^{\otimes q(n)}\Big| \mathcal{H}=H_0\right] 
\nonumber \\ 
&&	+		\Pr\left[ \hat{\mathcal{H}}_2=H_1, L_{q(n)}=1\Big| \mathcal{H}=H_0\right]
			\\	
			& \leq & \Pr\left[	 L_{q(n)}=0\Big|(W_1,\ldots, W_{q(n)}) = w_0^{\otimes q(n)}, \mathcal{H}=H_0\right] +\Pr\left[(W_1,\ldots, W_{q(n)}) = w_1^{\otimes q(n)}\Big| \mathcal{H}=H_0\right] 
\nonumber \\ 
&&	+		\Pr\left[ \hat{\mathcal{H}}_2=H_1\Big| \mathcal{H}=H_0\right]
			\\
				&\leq&
			\mu/3 
				+(\mu/3+\epsilon') +
				\mu/3= \epsilon ,
	\end{IEEEeqnarray}
where	the last inequality holds by the threshold chosen for the Neyman-Pearson test, by the properties of the typical set and the set $\mathcal{S}_n$, and because both the probability of channel decoding error and of wrong hypothesis testing vanish as $n \to \infty$, see for example \cite{SW18}.

Before analyzing the type-II error probability, we  notice  that $\hat{\mathcal H}=H_0$ is only possible when  $L_{q(n)}=1$ and $\hat{M}\neq 0$, in which case   $\hat{\mathcal H}_{2}=\hat{\mathcal H}$ and $\hat{M}_2=\hat{M}\geq 1$. Therefore, for the type-II error probability:
\begin{IEEEeqnarray}{rCl}
	\Pr\left[ \hat{\mathcal{H}}=H_0 \Big| \mathcal{H}=H_1 \right] 
	&=& \Pr\left[\hat{\mathcal{H}}=H_0 ,  L_{q(n)}=1, \hat{M}\neq 0\Big| \mathcal{H}=H_1 \right]\\
	 &=& \Pr\left[\hat{\mathcal{H}}_2=H_0 ,  L_{q(n)}=1, \hat{M}_2\neq 0 \Big| \mathcal{H}=H_1 \right]\\
	& \leq &  \Pr\left[\hat{\mathcal{H}}_2=H_0 \Big| \hat{M}_2\neq 0, \mathcal{H}=H_1 \right]\\
	& = &  \Pr\left[(S^n(\hat{M}_2),Y^n)\in\mathcal{T}_{\mu}^n(P_{SY}) \Big| \hat{M}_2\neq 0, \mathcal{H}=H_1 \right].
 	\end{IEEEeqnarray}
Under $H_1$, the observations $Y^n$ are i.i.d. according to $P_Y$ and  independent of $(S^n(\tilde{M}), \tilde{M})$, and thus by a conditional version of Sanov's theorem and continuity of the mutual information measure:
\begin{IEEEeqnarray}{rCl}	
\Pr\left[ 
(S^n(\hat{M}_2),Y^n)\in\mathcal{T}_{\mu}^n(P_{SY})
 \Big|  \hat{M}_2\neq 0,\mathcal{H}=H_1 \right]	&\leq & 2^{-n(I(S;Y)+\delta(\mu))},
	\end{IEEEeqnarray}
where $\delta(\mu)$ is a function that tends to 0 as $\mu \to 0$. Combining these last two inequalities, one obtains:
\begin{IEEEeqnarray}{rCl}
\Pr\left[ \hat{\mathcal{H}}=H_0 \Big| \mathcal{H}=H_1 \right]  &\leq & 2^{-n(I(S;Y)+\delta(\mu))}.
 	\end{IEEEeqnarray}
Taking $n\to \infty$ and $\mu\to 0$, it can be concluded that averaged over the random code construction the desired error exponent is achievable. By standard arguments it then follows that there exist deterministic codebooks achieving the desired exponents. 
\end{IEEEproof}

\section{Proof of Converse to Theorem~\ref{thm1}}\label{sec:converse}
Before proving the converse, we state a standard  auxiliary lemma commonly used for hypothesis testing  converses. \begin{lemma}\label{lem:weak_converse}
	Let $Q$ and $P$ be arbitrary pmfs over a discrete and finite set $\mathcal{Z}$ and $\mathcal{A}$ be a subset of $\mathcal{Z}$. Then, 
	\begin{equation}
	-\log Q(\mathcal{A})\leq \frac{1}{P(\mathcal{A})} ( D(P \|Q)+ 1).
	\end{equation}
\end{lemma}
\begin{IEEEproof}
	By the data processing inequality for KL-divergence: 
	\begin{IEEEeqnarray}{rCl}
		D(P \| Q) &\geq & P(\mathcal{A})  \log \frac{P(\mathcal{A}) }{Q(\mathcal{A}) } + (1-P(\mathcal{A}) ) \log \frac{(1-P(\mathcal{A})) }{(1-Q(\mathcal{A}) )} \\
		& = & -H_b( P(\mathcal{A})) - P(\mathcal{A})  \log Q(\mathcal{A})  -  (1-P(\mathcal{A}) ) \log (1-Q(\mathcal{A}) ).
	\end{IEEEeqnarray}
	Upper bounding $H_b( P(\mathcal{A}))$ by $1$ and $(1-P(\mathcal{A}) ) \log (1-Q(\mathcal{A}) )$ by 0, and rearranging terms yields the desired inequality. 
\end{IEEEproof}

We now prove the desired converse. Fix an achievable exponent $\theta<\theta^*_{\epsilon}(R)$ and a sequence of encoding and decision functions so that \eqref{type-I-def} and \eqref{type-II-cons} are satisfied. 
Further fix a blocklength $n>0$ and let $\M$ and $\hat{\mathcal{H}}$ be the bit-string message and the guess produced by the chosen encoding and decision functions for this given blocklength. Let then  $\mu, \eta$ be small positive numbers and define $\mathcal{B}_n(\eta)$ as a subset of $\mathcal{X}^n\times \mathcal{M}$:
\begin{IEEEeqnarray}{rCl}\label{eq:Bn}
	\mathcal{B}_n(\eta) \triangleq \left\{ (x^n,\m)\colon \Pr\big[\hat{\mathcal{H}}=H_0\big |X^n=x^n, \M=\m,\mathcal{H}=H_0\big]\geq \eta \right\}. 
\end{IEEEeqnarray}
By the constraint on the type-I error probability, \eqref{type-I-def},
\begin{IEEEeqnarray}{rCl}
	1-\epsilon &\leq & 
	\sum_{(x^n,\m)\in\mathcal{B}_n(\eta)}\Pr\big[\hat{\mathcal{H}}=H_0\Big|X^n=x^n,\M=\m,\mathcal{H}=H_0\big]\cdot P_{X^n\M}(x^n,\m) \nonumber \\
	&&+ \sum_{(x^n,\m)\in( \mathcal{Y}^n\times \set{M} )\backslash \mathcal{B}_n(\eta)}\Pr\big[\hat{\mathcal{H}}=H_0\Big |X^n=x^n,\M=\m,\mathcal{H}=H_0\big]\cdot P_{X^n\M}(x^n,\m) \IEEEeqnarraynumspace\\
	&\leq &P_{X^n\M}(\mathcal{B}_n(\eta))+\eta(1-P_{X^n\M}(\mathcal{B}_n(\eta))),
\end{IEEEeqnarray}	
and as a consequence: 
\begin{IEEEeqnarray}{rCl}
	P_{X^n\M}(\mathcal{B}_n(\eta))\geq \frac{1-\epsilon-\eta}{1-\eta}.\label{step1} 
\end{IEEEeqnarray}

We next define the subset $\mathcal{D}_n(\eta)$ of $\mathcal{X}\times \set{M}$:
\begin{align}\
\mathcal{D}_n(\eta)\triangleq \mathcal{B}_n(\eta) \cap ( \mathcal{T}_{\mu}^{n}(P_X)\times \set{M})
\end{align} 
By  \cite[Lemma 2.12]{Csiszarbook}:
\begin{IEEEeqnarray}{rCl}
	P_X^n(\mathcal{T}_{\mu}^{n}(P_X)) \geq 1-\frac{|\mathcal{X}|}{2\mu n},\label{step2}
\end{IEEEeqnarray}
which combined with 
\eqref{step1} and  the general identity $\Pr(A\cap B)\geq \Pr(A)+\Pr(B)-1$ implies:
\begin{IEEEeqnarray}{rCl}\label{eq:Deltan}
	P_{X^n\M}(\mathcal{D}_n(\eta)) \geq \frac{1-\epsilon-\eta}{1-\eta}-\frac{|\mathcal{X}|}{2\mu n}\triangleq \Delta_n.
\end{IEEEeqnarray}

Define  finally the random variables $(\tilde{\M},\tilde{X}^n,\tilde{Y}^n)$ as the restriction of the triple $(\M, X^n, Y^n)$ to  $(X^n,\M) \in \mathcal{D}_n(\eta)$. The probability distribution of the restricted triple  is then given by:
\begin{IEEEeqnarray}{rCl}
	P_{\tilde{\M}\tilde{X}^n\tilde{Y}^n}(\mathsf{m},x^n,y^n)\triangleq P_{XY}^n(x^n,y^n)\cdot \frac{\mathbbm{1}\left\{( x^n,\m)\in\mathcal{D}_n(\eta) \right\}}{\Pr(\mathcal{D}_n(\eta))}
	\IEEEeqnarraynumspace \label{tilde-def}
\end{IEEEeqnarray}
This implies in particular:
\begin{IEEEeqnarray}{rCl}
	P_{\tilde{X}^n}(x^n)&\leq& P_X^n(x^n)\cdot 
	\Delta_n^{-1},\label{step11}\\
	P_{\tilde{Y}^n}(y^n)&\leq& P_Y^n(y^n)\cdot 
	\Delta_n^{-1},\label{mar1}\\
	P_{\tilde{\M}}(\mathsf{m}) &\leq & P_{\M}(\mathsf{m})\cdot \Delta_n^{-1}
	\label{mar2}
\end{IEEEeqnarray}
and
\begin{IEEEeqnarray}{rCl}
	D\left(P_{\tilde{X}^n}\|P_X^n\right)\leq \log
	\Delta_n^{-1}. \label{KL-tilde}
\end{IEEEeqnarray}

We are now ready to provide  a lower bound on the expected rate and an upper bound on the  type-II error exponent with the desired single-letter correspondences in the asymptotic regimes where the blocklength grows to $\infty$ and the parameters $\mu,\eta \to 0$. 

\noindent\underline{\textit{Lower bound on the expected rate}}: Define the random variable $\tilde{L}\triangleq \len(\tilde{\M})$ and notice that by the rate constraint \eqref{L-def}:
\begin{IEEEeqnarray}{rCl}
	nR &\geq & \mathbb{E}\left[L\right]\\
	&=& \mathbb{E}\left[L|(X^n,\M)\in\mathcal{D}_n(\eta)\right]\cdot P_{X^n\M}(\mathcal{D}_n(\eta))+\mathbb{E}\left[L|X^n\notin\mathcal{D}_n(\eta)\right]\cdot (1-P_{X^n\M}(\mathcal{D}_n(\eta)))\\
	&\geq & \mathbb{E}\left[L|X^n\in\mathcal{D}_n(\eta)\right]\cdot P_{X^n\M}(\mathcal{D}_n(\eta))\\
	&= & \mathbb{E}\left[\tilde{L}\right]\cdot P_{X^n\M}(\mathcal{D}_n(\eta))\label{step60}\\
	&\geq & \mathbb{E}\left[\tilde{L}\right]\cdot \Delta_n,\label{step3}
\end{IEEEeqnarray}
where \eqref{step60} holds because $\tilde{\M}$ is obtained by restricting $\M$ to the event $(X^n ,\M)\in \mathcal{D}_n(\eta)$ and $\tilde{L}$ denotes the length of $\tilde{\M}$; and step~\eqref{step3} holds by the definition of $\Delta_n$ in \eqref{eq:Deltan}.

Now, since $\tilde{L}$ is  a function of $\tilde{\M}$, we  have:
\begin{IEEEeqnarray}{rCl}
	H(\tilde{\M}) &=& H(\tilde{\M},\tilde{L})\\
	&=& H(\tilde{\M}|\tilde{L})+H(\tilde{L})\\
	&=& \sum_{\ell} \Pr(\tilde{L}=\ell)H(\tilde{\M}|\tilde{L}=\ell)+H(\tilde{L})\\
	&\leq & \sum_{\ell} \Pr(\tilde{L}=\ell)\ell+H(\tilde{L})\\
	&=& \mathbb{E}[\tilde{L}]+H(\tilde{L})\label{step1002}\\
	&\leq & \frac{nR}{\Delta_n}+H(\tilde{L})\label{step4}\\
	&\leq &  \frac{nR}{\Delta_n}+\frac{nR}{\Delta_n}h_{\text{b}}\left(\frac{\Delta_n}{nR}\right)\label{step5}\\
	&=& \frac{nR}{\Delta_n}\left(1+h_{\text{b}}\left(\frac{\Delta_n}{nR}\right)\right).
	\label{step12}
\end{IEEEeqnarray}
Here, \eqref{step4} follows from \eqref{step3}; and \eqref{step5} holds because when $\mathbb{E}[\tilde{L}]\leq \frac{nR}{\Delta_n}$, then the entropy of $\tilde{L}$ can be at most that of a Geometric distribution with mean $\frac{nR}{\Delta_n}$, which is $\frac{nR}{\Delta_n}\cdot h_{\text{b}}\left(\frac{\Delta_n}{nR}\right)$.

On the other hand, we  can  lower bound $H(\tilde{\M})$ in the following way: 
\begin{IEEEeqnarray}{rCl}
	H(\tilde{\M}) &\geq & I(\tilde{\M};\tilde{X}^n)\label{eq1}\\
	&=& H(\tilde{X}^n)-H(\tilde{X}^n|\tilde{\M})\label{eq:step20_start}\\
	&=& -\sum_{x^n} P_{\tilde{X}^n}(x^n)\log P_{\tilde{X}^n}(x^n)-H(\tilde{X}^n|\tilde{\M})\\
	&\geq& -\sum_{x^n} P_{\tilde{X}^n}(x^n)\log P_{X^n}(x^n)+\log \Delta_n-H(\tilde{X}^n|\tilde{\M})\label{step6}\\
	&=& -\sum_{x^n} P_{\tilde{X}^n}(x^n)\sum_{t=1}^n\log P_{X}(x_t)+\log \Delta_n-H(\tilde{X}^n|\tilde{\M})\label{step7}\\
	&=& -\sum_{t=1}^n\sum_{x_t} P_{\tilde{X}_t}(x_t)\log P_{X}(x_t)+\log \Delta_n-H(\tilde{X}^n|\tilde{\M})\\
	&= &\sum_{t=1}^nH(\tilde{X}_t)+\sum_{t=1}^n D(P_{\tilde{X}_t}\|P_{X})+\log \Delta_n-H(\tilde{X}^n|\tilde{\M})\label{step8}\\
	&=&\sum_{t=1}^n\left[H(\tilde{X}_t)-H(\tilde{X}_t|\tilde{M},\tilde{X}^{t-1})\right]+\sum_{t=1}^n D(P_{\tilde{X}_t}\|P_{X})+\log \Delta_n\\
	&=&\sum_{t=1}^nI(\tilde{U}_t;\tilde{X}_t)+\sum_{t=1}^n D(P_{\tilde{X}_t}\|P_{X})+\log \Delta_n\label{step9}\\
	&=&nI(\tilde{U}_T;\tilde{X}_T|T)+\sum_{t=1}^n\;\sum_{x\in\mathcal{X}} P_{\tilde{X}_T|T=t}(x)\log \frac{P_{\tilde{X}_T|T=t}(x)}{P_{X}(x)}+\log\Delta_n\\
	&=& nI(\tilde{U}_T;\tilde{X}_T|T)+\sum_{t=1}^n\;\sum_{x\in\mathcal{X}} P_{\tilde{X}_T|T=t}(x)\log \frac{P_{\tilde{X}_T|T=t}(x)}{P_{\tilde{X}_T}(x)}+\sum_{t=1}^n\;\sum_{x\in\mathcal{X}} P_{\tilde{X}_T|T=t}(x)\log \frac{P_{\tilde{X}_T}(x)}{P_{X_t}(x)}+\log \Delta_n\nonumber\\\\
	&=& nI(\tilde{U}_T;\tilde{X}_T|T)+nI(\tilde{X}_T;T)+nD(P_{\tilde{X}_T}\|P_{X_T})+\log \Delta_n\label{step10}\\
	&\geq & nI(\tilde{U}_T,T;\tilde{X}_T)+\log \Delta_n\\
	&=&nI(\tilde U;\tilde{X}_T)+\log \Delta_n,\label{step20}
\end{IEEEeqnarray}
where 
\begin{itemize}
	\item \eqref{step6} holds by \eqref{step11};
	\item \eqref{step7} holds because $X^n$ is i.i.d. under $P_X^n$;
	\item \eqref{step9} holds by defining $\tilde{U}_t\triangleq (\tilde{\M},\tilde{X}^{t-1})$;
	\item \eqref{step10} holds because $T$ is uniformly chosen over $\{1,\ldots,n\}$;
	\item \eqref{step20} follows by defining $\tilde U\triangleq (\tilde{U}_T,T)$.
\end{itemize}
Combining \eqref{step12} and \eqref{step20}, we obtain:
\begin{IEEEeqnarray}{rCl}\label{eq:la}
	R\geq \frac{  I(\tilde U; \tilde{X}_T)+ \frac{1}{n} \log \Delta_n}{ 1+h_{\text{b}}\left(\frac{\Delta_n}{nR} \right)} \cdot \Delta_n,
\end{IEEEeqnarray} 
and conclude that in the limit $n \to \infty$ the rate $R$ needs to be lower bounded by the limit of the mutual information $I(\tilde U;\tilde{X})\frac{1-\eta-\epsilon}{1-\eta}$.
\bigskip 

\noindent\underline{\textit{Upper bound on the type-II error exponent}}:
For each string $\m\in\{0,1\}^\star$, define the following set:
\begin{IEEEeqnarray}{rCl}
	\mathcal{A}_n(\m)\triangleq \{y^n\colon (\m,y^n)\in\mathcal{A}_n \}, 
\end{IEEEeqnarray}
By definition of the set $\mathcal{D}_n(\eta)$:
\begin{IEEEeqnarray}{rCl}\label{eq:prob_eta}
	P_{Y|X}^n(\mathcal{A}_n(\m)|x^n)\geq \eta, \qquad  (x^n,\m) \in \mathcal{D}_n(\eta).
\end{IEEEeqnarray}
Let now $\{\ell_n\}_{n\geq 1}$ be a sequence  satisfying $\lim_{n\to\infty}\ell_n/\sqrt{n}=\infty$ and $\lim_{n\to\infty}\ell_n/n=0$, and define   for each $\m \in \set{M}$ the blown up region
\begin{IEEEeqnarray}{rCl}
	\hat{\mathcal{A}}_n^{\ell_n}(\m)\triangleq \left\{ \tilde{y}^n\colon \exists y^n\in\mathcal{A}_n(\m)\;\;\text{s.t.}\;\;d_{\text{H}}(\tilde{y}^n,y^n)\leq \ell_n \right\}.
\end{IEEEeqnarray}
By \eqref{eq:prob_eta} and the blowing-up lemma \cite[remark p. 446]{MartonBU}:
\begin{IEEEeqnarray}{rCl}
	P_{Y|X}^n\left(\hat{\mathcal{A}}_n^{\ell_n}(\m)\Big|x^n\right)\geq	1-\frac{\sqrt{n\ln 1/\eta}}{\ell_n}= 1-\lambda_n, \qquad  (x^n,\m) \in \mathcal{D}_n(\eta),\label{eq:dd}
\end{IEEEeqnarray}
where we defined $\lambda_n\triangleq \frac{\sqrt{n \ln 1/\eta}}{\ell_n}$. (Notice that $\lambda_n$ goes to zero as $n\to\infty$.)
Defining  the new acceptance region
\begin{IEEEeqnarray}{rCl}
	\hat{\mathcal{A}}_n^{\ell_n}\triangleq \bigcup_{m \in \set{M}} \{\m\}\times \hat{\mathcal{A}}_n^{\ell_n}(\m),
\end{IEEEeqnarray}
and taking expectation over \eqref{eq:dd}, we obtain:
\begin{IEEEeqnarray}{rCl}
	P_{\tilde{M}\tilde{Y}^n}(\hat{\mathcal{A}}_n^{\ell_n})=\sum_{(x^n,\m)\in\mathcal{D}_n(\eta)}P_{Y|X}^n(\hat{\mathcal{A}}_n^{\ell_n}(\m)|x^n)\cdot P_{\tilde X^n\tilde \M}(x^n,\m)\geq 1-\lambda_n.\label{typeI}
\end{IEEEeqnarray}

We next show that the probability of this new acceptance region under the product distribution $P_{\tilde{M}}P_{\tilde{Y}^n}$ is close (in terms of exponential decay rate) to the type-II error probability of our original hypothesis testing problem:
\begin{IEEEeqnarray}{rCl}
	P_{\tilde{M}}P_{\tilde{Y}^n}(\hat{\mathcal{A}}_n^{\ell_n})
	&\leq & P_{M}P_Y^n(\hat{\mathcal{A}}_n^{\ell_n})\cdot \Delta_n^{-2}\label{step63}\\
	&\leq & P_{M}P_Y^n(\mathcal{A}_n)\cdot e^{n h_{\textnormal{b}}(\ell_n/n)} \cdot|\mathcal{Y}|^{\ell_n}\cdot K_n^{\ell_n} \cdot \Delta_n^{-2}\label{step62}\\
	&=&\beta_n\cdot  e^{n h_{\textnormal{b}}(\ell_n/n)} \cdot|\mathcal{Y}|^{\ell_n}\cdot K_n^{\ell_n}  \cdot \Delta_n^{-2},\label{step64}
\end{IEEEeqnarray}
where we defined $K_n\triangleq  \min_{y:P_Y(y')>0} P_Y(y)$, and where  \eqref{step63} holds by \eqref{step11} and 
\eqref{step62}  by  \cite[see the Proof of Lemma 5.1]{Csiszarbook}.
Define $\delta_n\triangleq -\frac{2}{n}\log \Delta_n+\frac{\ell_n}{n}\log( K_n|\mathcal{Y}|) +h_{\textnormal{b}}(\ell_n/n)$ and notice that $\delta_n \to 0$ as  $n\to \infty$. We rewrite  \eqref{step64}  as  
\begin{IEEEeqnarray}{rCl}
	-\frac{1}{n}\log \beta_n &\leq & -\frac{1}{n}\log P_{\tilde{M}}P_{\tilde{Y}^n}(\hat{\mathcal{A}}_n^{\ell_n})+\delta_n \label{eq:step_overload1}\\
	&\leq &	\frac{1}{n(1-\lambda_n)}[D\left(P_{\tilde{\M}\tilde{Y}^n}\|P_{\tilde{\M}}P_{\tilde{Y}^n}\right)+1]+\delta_n\label{step37}\\
	&= & 	\frac{1}{n(1-\lambda_n)}\Big[ I(\tilde{\M};\tilde{Y}^n)+1\Big]+\delta_n\\
	&=&	 \frac{1}{n(1-\lambda_n)}\Big[ \sum_{t=1}^nI(\tilde{\M};\tilde{Y}_t|\tilde{Y}^{t-1})+1\Big]+\delta_n\\
	&\leq &	\frac{1}{n(1-\lambda_n)}\Big[\sum_{t=1}^nI(\tilde{\M},\tilde{Y}^{t-1};\tilde{Y}_t)+1\Big]+	\delta_n\\
	&\leq &	\frac{1}{n(1-\lambda_n)}\Big [\sum_{t=1}^nI(\underbrace{\tilde{\M},\tilde{X}^{t-1}}_{= \tilde{U}_{t}};\tilde{Y}_t)+1\Big]+\delta_n\label{step22}\\
	&=&	  \frac{1}{n(1-\lambda_n)} \Big[ \sum_{t=1}^nI(\tilde{U}_t;\tilde{Y}_t)+1\Big]+	\delta_n\\
	&=& \frac{1}{n(1-\lambda_n)} [I(\tilde{U}_T;\tilde{Y}_T|T)+1]+\delta_n\\
	&\leq & \frac{1}{1-\lambda_n}  [I(\underbrace{\tilde{U}_T,T}_{ =\tilde{U}};\tilde{Y}_T)	+1]+\delta_n\\
	&\leq & \frac{1}{1-\lambda_n}  [I(\tilde{U};\tilde{Y}_T)	+1]+\delta_n,\label{step23}
\end{IEEEeqnarray}
where
\begin{itemize}
	\item \eqref{step37} holds by Lemma~\ref{lem:weak_converse} and Inequality~\eqref{typeI};
	\item \eqref{step22} holds by the Markov chain $\tilde{Y}^{t-1}\to (\tilde{\M},\tilde{X}^{t-1})\to \tilde{Y}_t$.
\end{itemize}

The alphabet  of $\tilde{U}$  grows exponentially in $n$. However, by Charathéodory's theorem, for each blocklength $n$ there exists a random variable ${U}_n$ 
over an alphabet of size $|\mathcal{X}|+1$ and so that  the Markov chain $U_n\to \tilde{X}_T \to \tilde{Y}_T$  and the equalities $I(U_n;\tilde X_T) = I(\tilde{U};\tilde X_T)$ and $I(U_n;\tilde Y_T) = I(\tilde{U};\tilde Y_T)$ are satisfied. We can thus replace in \eqref{eq:la} and \eqref{step23}  the random variable $\tilde{U}$ by this new random variable $U_n$.

The proof is then concluded by  taking $n\to \infty$ and then $\mu,\eta\to 0$. In fact, recall that $\tilde{X}^n \in \mathcal{T}_{\mu/2}(P_X)$ and $\tilde{Y}^n$ is obtained by passing $\tilde{X}^n$ through the memoryless channel $P_{Y|X}$, which implies that as $n\to \infty$ and $\mu \to 0$ the distribution of $P_{\tilde{X}_T\tilde{Y}_T}$ tends to $P_{XY}$. By standard continuity considerations, the modified bounds \eqref{eq:la} and \eqref{step23} with $\tilde{U}$ replaced by $U_n$, and because all random variables have fixed and finite alphabet sizes, we can then conclude that
\begin{IEEEeqnarray}{rCl}
	\varlimsup_{ n \to \infty} -\frac{1}{n}\log\beta_n \leq  I(U;{Y})
\end{IEEEeqnarray}
for a random variable $U$ satisfying 
\begin{equation}
R \geq I(U;X) (1-\epsilon)
\end{equation}
and the Markov chain $U\to X \to Y$ and $(X,Y)\sim P_{XY}$.

This concludes the proof of the converse.

\section{Proof of Converse to Theorem~\ref{thm:noisy}}\label{sec:converse2}
Fix  an achievable exponent $\theta<\theta^*_{\textnormal{DMC},\epsilon}(\kappa)$ and a sequence of encoding  functions $\{ \Phi_1^{(n)}, \Phi_2^{(n)}, \ldots\}_{n\geq 1}$, stopping functions   $\{ e_1^{(n)}, e_2^{(n)}, \ldots\}_{n\geq 1}$, and acceptance/rejection regions $\{\mathcal{A}_n,\mathcal{R}_n\}_{n\geq 1}$ so that \eqref{eq:t1}--\eqref{stop-time} are satisfied.  
Further fix a large blocklength $n$, and let $\tau_n,W^{\tau_n},V^{\tau_n}$ be the stopping time, channel inputs and outputs as implied by these encoding and stopping  functions. Let $\mu , \eta$ be small positive real numbers 
and define 
\begin{equation}
\sigma\triangleq \ln(n)\cdot n
\end{equation}
and a new acceptance region $\mathcal{A}_{n}^{\text{new}}\subseteq \mathcal{A}_n$ which only contains  output sequences $v^\tau$ of length not exceeding $\sigma$:
\begin{IEEEeqnarray}{rCl}
	\mathcal{A}_{n}^{\text{new}}\triangleq \left\{ (v^{\tau},y^n)\in \mathcal{V}^\star\times  \mathcal{Y}^n\colon  (v^{\tau},y^n)\in \mathcal{A}_n \textnormal{ and } \tau\leq \sigma\right\}.\nonumber\\\label{new-A}
\end{IEEEeqnarray}

Define also the set 
\begin{IEEEeqnarray}{rCl}
	\mathcal{D}_{n}(\eta) &\triangleq & \Big\{(x^n,w^{\sigma})\colon \Pr\Big[(V^{\tau_n},Y^n)\in\mathcal{A}_{n}^{\text{new}}|  \mathcal{H}=H_0,X^n=x^n,W'^{\sigma}=w^{\sigma}\Big]\geq \eta \Big\}
	\cap\;\; \left(\mathcal{T}_{\mu}^n(P_X)\times \mathcal{W}^{\sigma}\right).
\end{IEEEeqnarray}
Notice that the set $\mathcal{D}_n(\eta_n)$ is defined in terms of the random variable $W'^\sigma$ but not $W^\sigma$ because the actual transmission duration might be shorter than $\sigma$, i.e. $\tau_n<\sigma$ is possible.

By standard arguments, we have
\begin{IEEEeqnarray}{rCl}
	\hspace{-0.5cm}1-\epsilon &\leq & P_{V^{\tau_n}Y^n}(\mathcal{A}_{n})\\[1ex]
	&=&\Pr[\tau_n\leq \sigma]\cdot P_{V^{\tau_n}Y^n}(\mathcal{A}_{n}|\tau_n\leq \sigma)+\Pr[\tau_n\geq \sigma]P_{V^{\tau_n}Y^n}(\mathcal{A}_{n}|\tau_n\geq \sigma)\\[1ex]
	&\leq &  P_{V^{\tau_n}Y^n}(\mathcal{A}_{n}^{\text{new}})+\frac{\mathbb{E}[\tau_n]}{\sigma}\label{step40}\\[1ex]
	&\leq & P_{V^{\tau_n}Y^n}(\mathcal{A}_{n}^{\text{new}})+\frac{\kappa+\eta}{\ln(n)}\label{step41}\\[1ex]
	&=& \sum_{x^n,w^{\sigma}}P_{X^nW^{'m}}(x^n,w^{\sigma})\cdot\sum_{(v^{\tau},y^n)\in\mathcal{A}_{n}^{\text{new}}} P_{V^{\tau_n}Y^n|X^nW^{'\sigma}}(v^{\tau},y^n|x^n,w^{\sigma})+\frac{\kappa+\eta}{\ln(n)}\nonumber\\\\[1ex]
	&=& \sum_{(x^n,w^{\sigma})\in\mathcal{D}_{n}(\eta)}P_{X^nW^{'m}}(x^n,w^{\sigma})\cdot\sum_{(v^{\tau},y^n)\in\mathcal{A}_{n}^{\text{new}}} P_{V^{\tau_n}Y^n|X^nW^{'\sigma}}(v^{\tau},y^n|x^n,w^{\sigma})\nonumber\\[1ex]
	&&+ \sum_{(x^n,w^{\sigma})\in\mathcal{D}_{n}^c(\eta)}P_{X^nW^{'\sigma}}(x^n,w^{\sigma})\cdot\sum_{(v^{\tau},y^n)\in\mathcal{A}_{n}^{\text{new}}} P_{V^{\tau}Y^n|X^nW^{'\sigma}}(v^{\tau},y^n|x^n,w^\sigma)+\frac{\kappa+\eta}{\ln(n)}\nonumber\\\\[1ex]
	&\leq &P_{X^nW^{'\sigma}}(\mathcal{D}_{n}(\eta))+(1-P_{X^nW^{'\sigma}}(\mathcal{D}_{n}(\eta))) \cdot\eta+\frac{\kappa+\eta}{\ln(n)},\nonumber\\
\end{IEEEeqnarray}
where 
\begin{itemize}
	\item \eqref{step40} follows from the definition of the new acceptance region $\mathcal{A}_n^{\text{new}}$ in \eqref{new-A} and from Markov's inequality;
	\item \eqref{step41} follows from \eqref{stop-time} and the definition  $\sigma=\ln(n)\cdot n$.
\end{itemize}

This implies
\begin{IEEEeqnarray}{rCl}
	P_{X^nW'^{\sigma}}(\mathcal{D}_{n}(\eta))\geq \frac{1-\epsilon-\eta-\frac{\kappa+\eta}{\ln(n)}}{1-\eta}-\frac{|\mathcal{X}|}{2\mu n}\triangleq \Delta_n.\label{Del-ineq}
\end{IEEEeqnarray}

Define  then the random tuple $(\tilde{X}^n,\tilde{Y}^n,\tilde{\tau}_n,	\tilde{W'}^{\sigma}, \tilde{W}^{\tilde \tau_n}, \tilde{V}^{\tilde{\tau}_n})$ as the restriction of the tuple $(X^n, Y^n,\tau_n,W^{'\sigma},W^{\tau_n},V^{\tau_n})$ to  $(X^n,W'^{\sigma}) \in \mathcal{D}_{n}(\eta)$. (Here we consider both sequences $W'^\sigma$ and  $W^{\tilde \tau_n}$ but the restriction is only on sequences $W^{' \sigma}$.) The restricted  pmf is  given by
\begin{IEEEeqnarray}{rCl}
	\	P_{\tilde{X}^n\tilde{Y}^n\tilde{\tau}_n\tilde{W}^{'\sigma}\tilde{W}^{\tilde{\tau}_n}\tilde{V}^{\tilde{\tau}_n}}(x^n,y^n,\tau,w^{\sigma},w^{\tau},v^{\tau}) &\triangleq &
	\frac{P_{X^nW^{'\sigma}}(x^n,w^{\sigma})}{P_{X^nW^{'\sigma}}(\mathcal{D}_{n}(\eta))} \cdot \mathbbm{1}\left\{ (x^n,w^{\sigma})\in\mathcal{D}_{n}(\eta) \right\}  \nonumber\\[0.5ex]
	&&\hspace{0.2cm}\cdot P_{Y|X}^n(y^n|x^n)\cdot P_{\tau_n{W}^{{\tau}_n}V^{\tau_n}|W^{'\sigma}X^n}(\tau,w^{\tau},v^\tau| w^\sigma,x^n), \IEEEeqnarraynumspace\label{tilde-defb}
\end{IEEEeqnarray}
and satisfies
\begin{IEEEeqnarray}{rCl}
	P_{\tilde{X}^n\tilde{W}^{\tilde{\tau}_n}}(x^n,w^{\tau})&\leq& P_{X^nW^{\tau_n}}(x^n,w^{\tau})\cdot 
	\Delta_n^{-1},\label{step11b}\\
	P_{\tilde{Y}^n}(y^n)&\leq& P_Y^n(y^n)\cdot 
	\Delta_n^{-1},\label{mar1b}\\
	P_{\tilde{V}^{\tilde{\tau}_n}}(v^{\tau})&\leq &P_{V^{\tau_n}}(v^{\tau})\cdot \Delta_n^{-1}.\label{mar2b}
\end{IEEEeqnarray}

%
%

\medskip
\underline{\textit{Communication constraint}:} Similarly to \eqref{step3}, we obtain: 
\begin{IEEEeqnarray}{rCl}
	\mathbb{E}[\tau_n]
	&\geq &\mathbb{E}[\tilde{\tau}_n]\cdot \Delta_n,
\end{IEEEeqnarray}
Since the original transmission durations $\{\tau_n\}_{n=1}^\infty$ have to satisfy \eqref{stop-time}, for arbitrary $\eta>0$ and all sufficiently large blocklengths $n$:
\begin{equation}\label{eq:tau1}
\mathbb{E}[{\tilde{\tau}}_n]   \leq \mathbb{E}[{{\tau}}_n]\Delta_n^{-1} \leq n(\kappa+\eta)\Delta_n^{-1},
\end{equation}

Following the same steps as in \eqref{eq:step20_start}--\eqref{step20} but where $\tilde{\M}$ is replaced by $\tilde{V}^{\tilde{\tau}_n}$, we obtain: 
\begin{IEEEeqnarray}{rCl}
	\hspace{-0.5cm}I(\tilde{V}^{\tilde{\tau}_n};\tilde{X}^n)
	&\geq&nI(\tilde U;\tilde{X}_T)+\log \Delta_n,\label{step20b}
\end{IEEEeqnarray}
where here $\tilde U$ is defined as $I(\tilde{V}^{\tilde{\tau}_n},\tilde{X}^{T-1},T)$ for $T$ uniformly distributed over $\{1,\ldots,n\}$  independent of $(\tilde{V}^{\tilde{\tau}_n},\tilde{X}^{n}, \tilde{Y}^n)$.

In the following, we upper bound  $I(\tilde{V}^{\tilde{\tau}_n};\tilde{X}^n)$ by $n$ times the capacity $C$ of the DMC $\Gamma_{V|W}$ plus some additive terms that vanish in the asymptotic regimes $n\to\infty$ and $\eta,\mu \to 0$. Define for $i=1,2,\ldots$ the random variables 
$\tilde{L}_i\triangleq \mathbbm{1}\left\{ \tilde{\tau}_n\geq i \right\}$ and 
\begin{IEEEeqnarray}{rCl}
	\hat{V}_i\eqdef\left\{ \begin{array}{ll} \tilde{V}_i & \tilde{\tau}_n\geq i\\0 & \tilde{\tau}_n<i\end{array} \right.\label{start}
\end{IEEEeqnarray}
Notice that we can write $I(\tilde{V}^{\tilde{\tau}_n};\tilde{X}^n)$ as:
\begin{IEEEeqnarray}{rCl}
	I(\tilde{V}^{\tilde{\tau}_n};\tilde{X}^n) \; &=& I(\tilde{L}^{\infty},\hat{V}^{\infty};\tilde{X}^n)\label{step502}\\
	&=& \sum_{i=1}^{\infty} I(\tilde{L}_i,\hat{V}_i;\tilde{X}^n|\tilde{L}^{i-1},\hat{V}^{i-1})\\
	&= & \sum_{i=1}^{\infty} I(\tilde{L}_i;\tilde{X}^n|\tilde{L}^{i-1},\hat{V}^{i-1})+\sum_{i=1}^{\infty} I(\hat{V}_i;\tilde{X}^n|\tilde{L}^{i},\hat{V}^{i-1})\nonumber\\\\
	&= & \sum_{i=1}^{\infty} I(\tilde{L}_i;\tilde{X}^n|\tilde{L}^{i-1},\hat{V}^{i-1})+\sum_{i=1}^{\infty} I(\tilde{V}_i;\tilde{X}^n|\tilde{L}_i=1,\tilde{L}^{i-1},\tilde{V}^{i-1})\cdot \Pr[\tilde{L}_i=1]\label{step501}\\
	&\leq& \sum_{i=1}^{\infty} H(\tilde{L}_i|\tilde{L}^{i-1})+ \sum_{i=1}^{\infty}H(\tilde{V}_i)\cdot \Pr[\tilde{L}_i=1]-\sum_{i=1}^{\infty} H(\tilde{V}_i|\tilde{L}_i=1,\tilde{L}^{i-1},\tilde{W}_i,\tilde{V}^{i-1},\tilde{X}^n)\cdot \Pr[\tilde{L}_i=1]\nonumber\\\\
	&=& H(\tilde{L}^{\infty})+ \sum_{i=1}^{\infty} \left( H(\tilde{V}_i)-  H(\tilde{V}_i|\tilde{W}_i)\right)\cdot \Pr[\tilde{L}_i=1]\\
	&=& H(\tilde{L}^{\infty})+\sum_{i=1}^{\infty} I(\tilde{V}_i;\tilde{W}_i)\cdot \Pr[\tilde{L}_i=1]\label{step504}\\
	&\leq & H(\tilde{L}^{\infty})+C\cdot \sum_{i=1}^{\infty} \Pr[\tilde{L}_i=1]\label{step505}\\
	&\leq & H(\tilde{\tau}_n)+C\cdot \sum_{i=1}^{\infty}\Pr[\tilde{\tau}_n\geq i]\label{step500} \\
	&= & H(\tilde{\tau}_n)+C\cdot \mathbb{E}[\tilde{\tau}_n]\label{step3002} \\
	&\leq & \frac{n(\kappa+\eta)}{\Delta_n}\cdot h_{\text{b}}\left(\frac{\Delta_n}{n(\kappa+\eta)}\right)+nC ( \kappa +\eta)\Delta_n^{-1},\label{step3004}
\end{IEEEeqnarray}
where 
\begin{itemize}
	\item \eqref{step502} holds  because there is a bijective function from $(\tilde{L}^{\infty},\hat{V}^{\infty})$  to $\tilde{V}^{\tilde{\tau}_n}$;
	\item \eqref{step501} holds  because when $\tilde{L}_i=0$ then  $\hat{V}_i$ is deterministic and when $\tilde{L}_i=1$ then  $\hat{V}_i=\tilde{V}_i$;	\item \eqref{step504} holds  because when $\tilde{L}_i=1$ the Markov chain $\tilde{V}_i\to\tilde{W}_i\to (\tilde{L}^{i-1},\tilde{V}^{i-1},\tilde{X}^n)$ holds;		\item \eqref{step505} holds because $P_{\tilde{V}_i|\tilde{W}_i}=\Gamma_{V_i|W_i}$ and thus the mutual information term $I(\tilde{V}_i;\tilde{W}_i)$ is upper bounded by the  capacity $C$ of the channel;
	\item  \eqref{step500} holds because  there exists a bijective function from $\tilde{\tau}_n$  to $\tilde{L}^{\infty}$ and by the definition of $\tilde{L}_i$.
	and \item\eqref{step3004} holds  only for sufficiently large values of $n$, by
	\eqref{eq:tau1} and because when  $\mathbb{E}[\tilde{\tau_n}]\leq \frac{n(\kappa+\eta)}{\Delta_n}$, then the entropy of $\tilde{\tau}_n$ can be at most that of a Geometric distribution with mean $\frac{n(\kappa+\eta)}{\Delta_n}$, which is $\frac{n(\kappa+\eta)}{\Delta_n}\cdot h_{\text{b}}\left(\frac{\Delta_n}{n(\kappa+\eta)}\right)$. 
\end{itemize}

Combining \eqref{step20} and \eqref{step3004}, we  conclude that for all sufficiently large values of $n$:
\begin{IEEEeqnarray}{rCl}
	(\kappa +\eta)C   &\geq &  I(\tilde U;\tilde{X}_T)\cdot \Delta_n+\frac{\Delta_n}{n} \log \Delta_n - (\kappa+\eta)\cdot h_{\text{b}}\left(\frac{\Delta_n}{n(\kappa+\eta)}\right),\label{eq:part4}
\end{IEEEeqnarray}
and in particular, $(\kappa+\eta)\frac{1-\eta}{1-\eta -\epsilon}C$ upper bounds the limit of the mutual information $I(\tilde{U};\tilde{X}_T)$ as $n\to \infty$.

\medskip
\underline{\textit{Upper bounding the type-II error exponent}}:

By definition,
\begin{IEEEeqnarray}{rCl}
	P_{\tilde{V}^{\tilde{\tau}_n}\tilde{Y}^n|\tilde{X}^n\tilde{W}^{'\sigma}}(\mathcal{A}_{n}^{\textnormal{new}}|x^n,w^{\sigma}) \geq \eta, \quad \forall (x^n,w^{\sigma})\in\mathcal{D}_n(\eta).
\end{IEEEeqnarray}
We now expand the region $\mathcal{A}_{n}^{\textnormal{new}}$ to a subset of $\mathcal{V}^\sigma\times \mathcal{Y}^n$, i.e., we expand all channel output sequences to be of same length $\sigma$:
\begin{IEEEeqnarray}{rCl}
	\mathcal{A}_{n}^{\text{exp}} &\triangleq& \big\{ (v^{\sigma},y^n)\in \mathcal{V}^\sigma \times  \mathcal{Y}^n\colon   \exists (\tilde{v}^{\tau},y^n)\in \mathcal{A}_n   \textnormal{ and } \bar{v}^{\sigma-\tau} \colon  v^\sigma=(\tilde{v}^{\tau}, \bar{v}^{\sigma-\tau}) \big\}.\label{exp-A}\IEEEeqnarraynumspace \end{IEEEeqnarray}	
Similarly, let $\tilde V'^\sigma=(\tilde V_1',\ldots, \tilde V_\sigma')$ be outputs of the DMC  $\Gamma_{V|W}$ for inputs $\tilde W'^\sigma$, and in particular $V'_k=\tilde{V}_k$ with probability 1 when $k\leq\tilde{\tau}_n$.  Then, 
\begin{IEEEeqnarray}{rCl}
	P_{\tilde V'^\sigma\tilde{Y}^n|\tilde{X}^n\tilde{W}^{'\sigma}}( \mathcal{A}_n^{\textnormal{exp}}|x^n,w^\sigma) 
	& = & 	P_{\tilde{V}^{\tilde{\tau}_n}\tilde{Y}^n|\tilde{X}^n\tilde{W}^{'\sigma}}(\mathcal{A}_{n}^{\textnormal{new}}|x^n,w^{\sigma}) \\
	& \geq & \eta.
\end{IEEEeqnarray}
By the blowing-up lemma \cite[remark p. 446]{MartonBU}, 
\begin{IEEEeqnarray}{rCl}
	P_{\tilde V'^\sigma\tilde{Y}^n|\tilde{X}^n\tilde{W}^{'m}}( \hat{\mathcal{A}}_n^{\textnormal{exp},\ell_n}|x^n,w^\sigma) 
	& \geq & 1- \frac{\sqrt{ (n+\sigma) \ln (1/\eta)}}{\ell_n}= 1- \nu_n,
\end{IEEEeqnarray}
where we defined  $\nu_n\eqdef \frac{\sqrt{ (n+\sigma) \ln(1/\eta)}}{\ell_n}$ and the blown up region
\begin{IEEEeqnarray}{rCl}
	\hat{\mathcal{A}}_n^{\textnormal{exp},\ell_n}& \triangleq& \{ (\tilde{v}^\sigma, \tilde{y}^n)\colon \exists (v^\sigma,y^n)\in \mathcal{A}_n^{\textnormal{exp}} \textnormal{ s.t. }\; d_{\text{H}}(\tilde{v}^\sigma,v^\sigma)+ d_{\text{H}}(\tilde{y}^n,y^n)\leq \ell_n\}.
\end{IEEEeqnarray}
Averaging over the sequences $(x^n,w^\sigma) \in \mathcal{D}_n$ we obtain:
\begin{IEEEeqnarray}{rCl}
	P_{\tilde V'^\sigma\tilde{Y}^n}( \hat{\mathcal{A}}_n^{\textnormal{exp},\ell_n}) 
	& \geq & 1- \nu_n\label{eq:bound_nu}
\end{IEEEeqnarray}
Since  $\hat{\mathcal{A}}_n^{\textnormal{exp},\ell_n}$ is 
the expanded region of  $\hat{\mathcal{A}}_n^{\textnormal{new},\ell_n}$:
\begin{IEEEeqnarray}{rCl}
	P_{\tilde{V}^{\tilde{\tau}_n}\tilde{Y}^n}(\hat{\mathcal{A}}_{n}^{\textnormal{new},\ell_n}) 
	& = & 	
	P_{\tilde V'^m\tilde{Y}^n}( \hat{\mathcal{A}}_n^{\textnormal{exp},\ell_n})\\
	& \geq & 1- \nu_n.\label{step51}
\end{IEEEeqnarray}

Notice next:\begin{IEEEeqnarray}{rCl}
	P_{\tilde{V}^{\tilde{\tau}_n}} P_{\tilde Y^n}(\hat{\mathcal{A}}_{n}^{\textnormal{new},\ell_n}) &\leq& P_{V^{\tau_n}} P_{Y}^n(\hat{\mathcal{A}}_{n}^{\textnormal{new},\ell_n})\cdot \Delta_n^{-2}\label{step1c}\\
	&\leq & P_{V^{\tau_n}}P_{Y}^n(\mathcal{A}_{n}^{\textnormal{new}})\cdot K_n^{\ell_n}\cdot \Delta_n^{-2}\label{step2c} \IEEEeqnarraynumspace \\
	&\leq& \beta_n\cdot K_n^{\ell_n}\cdot \Delta_n^{-2},\label{eq:step3}
\end{IEEEeqnarray} 
where 
\begin{IEEEeqnarray}{rCl}
	K_n\triangleq \frac{n e}{\ell_n}pq|\mathcal{Y}||\mathcal{V}|
\end{IEEEeqnarray}
and 
\begin{IEEEeqnarray}{rCl}
	p&\triangleq &  \max_{y,y'\colon P_Y(y')>0}\frac{P_Y(y)}{P_Y(y')}\\
	q&\triangleq &  \max_{w,v,v':\Gamma_{V|W}(v'|w)>0}  \frac{\Gamma_{V|W}(v|w)}{\Gamma_{V|W}(v'|w)}. 
\end{IEEEeqnarray}

Here, \eqref{step1c} holds by {\eqref{mar1b}--\eqref{mar2b}} and for step~\eqref{step2c} see \cite[Proof of Lemma 5.1]{Csiszarbook}. Step~\eqref{eq:step3} holds because the original acceptance region includes the new region, $\mathcal{A}_n\supseteq \mathcal{A}_n^{\textnormal{new}}$.

We use \eqref{eq:step3} to bound the  type-II error exponent of the original test:
\begin{IEEEeqnarray}{rCl}
	-\frac{1}{n}\log\beta_n 
	&\leq & -\frac{1}{n}\log P_{\tilde{V}^{\tilde{\tau}_n}}P_{\tilde Y^n}(\hat{\mathcal{A}}_{n}^{\textnormal{new},\ell_n})-\frac{2}{n}\log\Delta_n+\frac{\ell_n}{n}\log K_n\label{step24}\IEEEeqnarraynumspace\\
	&\leq&\frac{1}{n(1- \nu_n)} \left( D(P_{\tilde{V}^{\tilde{\tau}_n}\tilde{Y}^n}\|P_{\tilde{V}^{\tilde{\tau}_n}} P_{\tilde Y^n})+ 1\right)) - \frac{2}{n}\log\Delta_n+\frac{\ell_n}{n}\log K_n \label{eq:part1}
\end{IEEEeqnarray}
where the second inequality  holds by Lemma~\ref{lem:weak_converse} stated at the beginning of Appendix~\ref{sec:converse} and by Inequality~\eqref{eq:bound_nu}.

We continue to single-letterize  the divergence term:
%
\begin{IEEEeqnarray}{rCl}
	\frac{1}{n}	D(P_{\tilde{V}^{\tilde{\tau}_n}\tilde{Y}^n}\|P_{\tilde{V}^{\tilde{\tau}_n}} P_{\tilde Y^n})
	&= & 	 \frac{1}{n}I(\tilde{V}^{\tilde{\tau}_n};\tilde{Y}^n)\\
	&=&	 \frac{1}{n} \sum_{t=1}^nI(\tilde{V}^{\tilde{\tau}_n};\tilde{Y}_t|\tilde{Y}^{t-1})\\
	&\leq &	 \frac{1}{n}\sum_{t=1}^nI(\tilde{V}^{\tilde{\tau}_n},\tilde{Y}^{t-1};\tilde{Y}_t)\\
	&\leq &	 \frac{1}{n}\sum_{t=1}^nI(\tilde{V}^{\tilde{\tau}_n},\tilde{X}^{t-1};\tilde{Y}_t)\label{step22b}\\
	&=&	 \frac{1}{n} \sum_{t=1}^nI(\tilde{U}_t;\tilde{Y}_t)\\
	&=& I(\tilde{U}_T;\tilde{Y}_T|T)\\
	&\leq & I(\tilde{U}_T,T;\tilde{Y}_T)\\
	&=& I(\tilde U;\tilde{Y}_T),\label{step23b}	
\end{IEEEeqnarray}
where
\eqref{step22b} holds by the Markov chain $\tilde{Y}^{t-1}\to (\tilde{V}^{\tilde{\tau}_n},\tilde{X}^{t-1})\to \tilde{Y}_t$.

Combining \eqref{eq:part1}  with \eqref{step23}, we obtain:
\begin{IEEEeqnarray}{rCl}
	-\frac{1}{n}\log\beta_n &\leq & \frac{1}{1- \nu_n} \left( I(\tilde U;\tilde{Y}_T) +\frac{1}{n}\right) - \frac{2}{n}\log\Delta_n+\frac{\ell_n}{n}\log K_n.\label{eq:part5}
\end{IEEEeqnarray}
When $n\to \infty$, then  $\nu_n\to 0$, $\frac{2}{n}\log\Delta_n\to 0$, and $\frac{\ell_n}{n}\log K_n\to 0$. So, the asymptotic type-II error exponent is upper bounded by the limit of $I(\tilde U;\tilde{Y}_T)$ as $n\to \infty$.  

We  analyze this limit. To this end, we notice that  by Charathéodory's theorem, for each blocklength $n$ there exists a random variable ${U}_n$ 
over an alphabet of size $|\mathcal{X}|+1$ and satisfying the Markov chain $U_n\to \tilde{X}_T \to \tilde{Y}_T$  and the equalities $I(U_n;\tilde X_T) = I(\tilde{U}_T;\tilde X_T)$ and $I(U_n;\tilde Y_T) = I(\tilde{U}_T;\tilde Y_T)$. We can thus replace in \eqref{eq:la} and \eqref{step23}  the random variable $\tilde{U}$ by this new random variable $U_n$.

The proof is then concluded by  taking $n\to \infty$ and then $\mu,\eta\to 0$. In fact, recall that  $\tilde{X}^n \in \mathcal{T}_{\mu}(P_X)$ and $\tilde{Y}^n$ is obtained by passing $\tilde{X}^n$ through the channel $P_{Y|X}$, which implies that as $n\to \infty$ and $\mu \to 0$ the distribution of $P_{\tilde{X}_T\tilde{Y}_T}$ tends to $P_{XY}$. By standard continuity considerations, the modified bounds \eqref{eq:part4} and \eqref{eq:part5} with $\tilde{U}$ replaced by $U_n$, and because all random variables have fixed and finite alphabet sizes, we can then conclude that
\begin{IEEEeqnarray}{rCl}
	\varlimsup_{ n \to \infty} -\frac{1}{n}\log\beta_n \leq  I(U;{Y})
\end{IEEEeqnarray}
for a random variable $U$ satisfying 
\begin{equation}
\frac{\kappa \cdot C}{1-\epsilon} \geq I(U;X) 
\end{equation}
and the Markov chain $U\to X \to Y$ and $(X,Y)\sim P_{XY}$.


\section{Conclusion and Remarks}\label{sec:con}

We established the optimal type-II error exponent of a distributed testing against independence problem under a constraint on the probability of type-I error and on the expected communication rate. This result can be seen as a variable-length coding version of the well-known result by Ahlswede and  Csisz\'ar \cite{Ahlswede} which holds under a maximum rate-constraint. Interestingly,  when the type-I error probability is constrained to be at most $\epsilon\in (0,1)$, then the optimal type-II error exponent under an expected rate constraint $R$ coincides with the optimal type- II error exponent under a maximum rate constraint $(1-\epsilon)R$. Thus, unlike in the scenario with a maximum rate constraint, here a strong converse does not hold, because the optimal type-II error exponent depends on the allowed type-I error probability $\epsilon$.

We also considered testing against independence over a DMC with variable-length coding and stop feedback. As we show, the optimal type-II error exponent depends on the DMC transition law only through its capacity. More specifically, under a type-I error probability constraint $\epsilon \in (0,1)$, the optimal type-II error exponent with variable-length coding over a DMC with capacity $C$ coincides with the optimal  type-II error exponent under fixed-length coding over a DMC with capacity $C/(1-\epsilon)$. Thus, a strong converse result does not hold for this setup, neither.

The paper considered setups where the marginal distributions are the same under both hypotheses. The presented results hold also when this assumption is relaxed, the important assumption is the independence of the sources under the alternative hypothesis $H_1$. An interesting future direction is to investigate whether also this assumption can be relaxed and similar results apply also for testing against conditional  independence. 

\section*{Acknowledgements}
S. Salehkalaibar and M. Wigger acknowledge  funding support from the ERC under grant  agreement 715111.
\bibliographystyle{IEEEtran}
\bibliography{references}

\appendices

\end{document}